%% file: Applying_a_Riesz_projection_based_contour_integral_eigenvalue_solver_to_compute_resonance_modes_of_a_VCSEL.tex
\documentclass[]{spie}  
\pdfoutput=1

\usepackage{amsmath,amsfonts,amssymb}
\usepackage{graphicx}
\usepackage[colorlinks=true, allcolors=blue]{hyperref}
\usepackage{multirow}

 \usepackage{tikz}
 \usetikzlibrary{datavisualization}
 \usetikzlibrary{plotmarks}
 \usetikzlibrary{plotmarks}
 \usetikzlibrary{graphs}
 \usetikzlibrary{shapes.geometric}
 \usetikzlibrary{arrows.meta}
 \usepackage{pgfplots}
 \pgfplotsset{compat=newest}
 \pgfplotsset{plot coordinates/math parser=false}
 \newlength\figureheight
 \newlength\figurewidth
 \usetikzlibrary{arrows}
 
 \usepackage{xcolor}

\usepackage{pgfplots}
\pgfplotsset{compat=newest}
\pgfplotsset{plot coordinates/math parser=false}
\usetikzlibrary{arrows}
\usepackage{subfig}

\title{Applying a Riesz-projection-based contour integral eigenvalue solver to compute resonance modes of a VCSEL}

\author[a,b]{Lilli Kuen}
\author[b]{Fridtjof Betz}
\author[b]{Felix Binkowski}
\author[a]{Philipp-Immanuel Schneider}
\author[a]{Martin Hammerschmidt}
\author[c]{Niels Heermeier}
\author[c]{Sven Rodt}
\author[c]{Stephan Reitzenstein}
\author[a,b]{Sven Burger}
\affil[a]{JCMwave GmbH, Bolivarallee 22, 14050 Berlin, Germany}
\affil[b]{Zuse Institute Berlin, Takustraße 7, 14195 Berlin, Germany}
\affil[c]{Institute of Solid State Physics, Technische Universität Berlin, 10623 Berlin, Germany}

\begin{document} 
\maketitle

\begin{abstract}
We investigate the calculation of resonance modes of a VCSEL with a Riesz projection eigenvalue solver. 
The eigenvalue solver is based on the principle of contour integration where for the solution of 
scattering problems physical right sides are used. 
Here, it is investigated how numerical parameters impact the performance of the method, 
where we focus on the computation of the fundamental VCSEL mode. 
\end{abstract}

\keywords{Riesz projection, eigenvalue solver, contour integration, quasi-normal mode, VCSEL}

 \noindent \textit{This paper will be published in Proc. SPIE Vol.}  \textbf{12575} \textit{(2023) 125750J (Integrated Optics: Design, Devices, Systems and Applications VII, DOI: 10.1117/12.2665490) and is made available as an electronic preprint with permission of SPIE. One
print or electronic copy may be made for personal use only. Systematic or multiple reproduction, distribution
to multiple locations via electronic or other means, duplication of any material in this paper for a fee or for
commercial purposes, or modification of the content of the paper are prohibited.}

\section{Introduction}
\label{sec:Introduction} 
Vertical-cavity surface-emitting lasers (VCSELs) are widely used light sources \cite{Liu:19}, 
for example, in communication \cite{gkebski202030, Moughames:20} and sensing \cite{moench2016vcsel,Haghighi2021} 
technologies. 
In the field of deep neural networks, coherent VCSELs can be used to build optical neuronal networks \cite{chen2022deep}.
Numerical simulation plays an important role in the design and optimization of these semiconductor lasers \cite{Bienstman,gkebski202030}.  The device dimension with a diameter of $30\,\mu \text{m}$ is large compared to the emission wavelength of $\lambda = 980\,\text{nm}$, which implies a relatively large numerical effort for 3D optical simulations. 
In the numerical investigation of VCSELs, the computation of eigenmodes and especially of the fundamental mode is of essential interest. 
The intensity maximum of the field distribution of resonant eigensolutions is inside the VCSEL cavity. 
The field decays inside the mirrors and is transmitted to the exterior mainly via the upper mirror. 
The fundamental mode field distribution in the radial direction is approximately of Gaussian shape. 
Its identification and computation are complex, due to the densely populated spectrum. 
A widely used eigenvalue solver is, e.g., the Arnoldi method \cite{saad2011numerical}. 
If one uses this to determine the fundamental mode of the VCSEL, a larger number of eigenvalues has to be computed. 
Typically, the fundamental mode can be found via the selection of the corresponding field distribution.
An alternative approach for solving the eigenproblem is the use of contour 
integration~\cite{Asakura_JSIAM_2009, beyn2012integral, Gavin_JCompPhy_2018}.
Contour integral methods compute eigenvalues in a chosen region in the complex plane, 
and they are essentially based on solving linear systems of equations.

Here, we apply a Riesz projection eigenvalue solver \cite{binkowski2020riesz,betz2021rpexpand} 
to compute the eigenvalues of resonance modes in a VCSEL. 
This method is a specific contour integration method. 
In the algorithm, a contour is placed within the complex plane of the eigenvalues. 
Then, scattered fields are computed at discrete frequencies along the contour by solving  
time-harmonic scattering problems with source terms at the corresponding complex frequencies. 
Finally, from these scattering simulations, enclosed eigenvalues and corresponding eigenvectors are obtained.  
The method differs from conventional contour integration methods, such as the algorithm proposed by Beyn \cite{beyn2012integral}, which uses random vectors as right-hand sides and computes all eigenvalues within the integration contour. 
Riesz projection eigenvalue solvers instead use physical sources and are 
capable to access only the physical eigenmodes of interest, e.g., 
eigenmodes with a specific polarization \cite{binkowski2020riesz}. 
Partial derivatives of the eigenvalues, that can be used for optimization, 
can be obtained with negligible or small additional effort \cite{binkowski2022computation}. 

We analyze the performance of the Riesz projection eigenvalue solver for the example of 
computing the fundamental mode of a VCSEL model. 
We investigate different setups for the contour enclosing one or several eigenvalues. 
For the excitation, three different types of sources are tested: 
plane waves, dipole sources, and a constant current source defined in a volume. 
For all setups, the convergence for each of the excitation sources is analyzed, 
as well as the corresponding eigenmodes. 
The results are verified by comparison to results from an Arnoldi method eigenvalue solver. 

\begin{figure}
    \centering
    \subfloat[]{\input{Bilder/SchematischeDarstellung.tex} \label{fig:FigVCSELSchemea}} \hspace{1cm}
    \subfloat[]{\input{Bilder/SchematischeDarstellung_NumerischesSetup.tex} \label{fig:FigVCSELSchemeb}}
    \caption{
    (a) Cross section of a cylindrical VCSEL with bottom and top distributed Bragg reflector (DBR). 
    Inside the cavity are several layers of active material in the form of quantum wells (MQW). 
    The mode is limited by an oxide aperture at the boundaries to the respective DBR mirrors.  
    (b) Setup for numerical simulation in $x$-$y$-coordinates with rotational symmetry (axis of rotation: $y$). 
    Transparent boundaries are indicated with a blue frame.  
    The different sources, (1) plane wave, (2) dipole, and (3) constant current, are indicated in the sketch. 
    Both sketches are not to scale and the numbers of layers (DBR and MQW) are simplified.
    }
    \label{fig:FigVCSELScheme}
\end{figure}
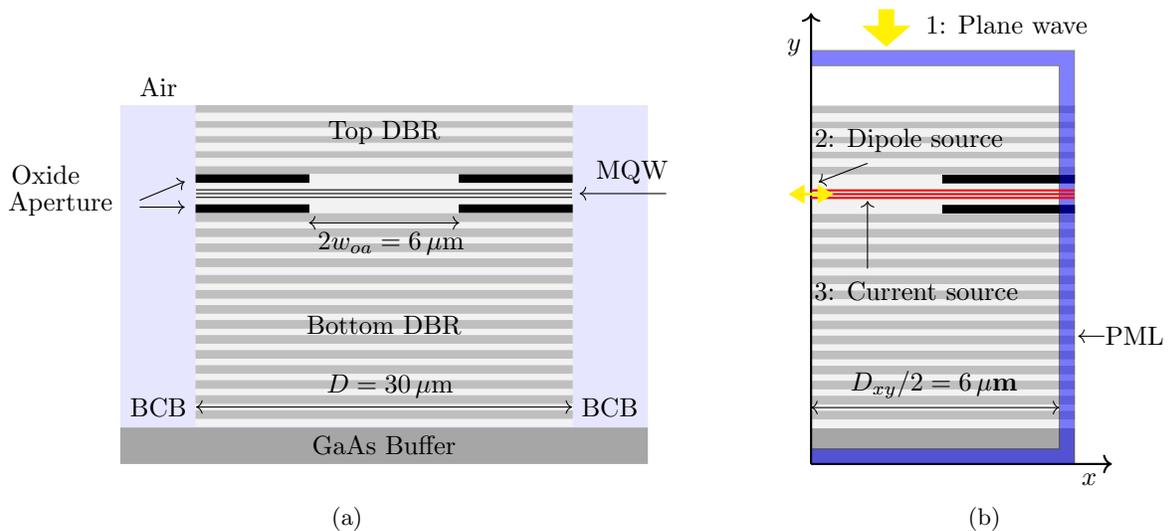

\section{Background} 
\label{sec:BasicInformations}
The electric field distributions, $\tilde{\mathbf{E}}(\mathbf{r})$, also called quasi-normal modes, QNMs, 
and the complex-valued eigenvalues, 
$\tilde{\omega} \in \mathbb{C}$, 
corresponding to resonances of a VCSEL can be obtained 
by solving the source-free time-harmonic Maxwell equation in second-order form,
\begin{equation}
    \nabla \times\mu(\mathbf{r},\tilde{\omega})^{-1} \nabla \times \tilde{\mathbf{E}}(\mathbf{r},\tilde{\omega}) - \tilde{\omega}^2 \varepsilon(\mathbf{r},\tilde{\omega})\tilde{\mathbf{E}}(\mathbf{r},\tilde{\omega}) = 0,
    \label{eq:TimeHarmonicMaxwell}
\end{equation}
with transparent boundary conditions, and with permeability, $\mu$, permittivity, $\varepsilon$, 
$\varepsilon=\varepsilon_r \varepsilon_0$, relative permittivity $\varepsilon_r$, and vacuum permittivity, $\varepsilon_0$.
For optical frequencies, typically, $\mu=\mu_0$, with the vacuum permittivity $\mu_0$. 
To solve Eq.~\eqref{eq:TimeHarmonicMaxwell} numerically, various methods can be used. 
Section~\ref{subsec:RieszProjection} reviews the background of the Riesz projection eigenvalue solver
which will be used throughout this work, in an implementation in the software package RPExpand \cite{betz2021rpexpand},
with the contained interface to the finite element method (FEM) solver JCMsuite \cite{JCMsuite,pomplun2007adaptive}.
For obtaining reference solutions we solve Eq.~\eqref{eq:TimeHarmonicMaxwell}
using an alternative method, the Arnoldi method \cite{saad2011numerical} implemented solver JCMsuite.

\subsection{Contour integration eigenvalue solver} 
\label{subsec:RieszProjection}
We review the theoretical background for the Riesz projection eigenvalue solver
\cite{binkowski2020riesz}, 
where essentially scattering problems are solved for obtaining eigenmodes and eigenvalues. 
Scattering solutions $\mathbf{E}(\mathbf{r},\omega_0 \in \mathbb{R})$ are solutions to 
Eq.~\eqref{eq:TimeHarmonicMaxwell}, but with a nonzero right-hand side describing a light source driving at the real frequency $\omega_0$.
With the Riesz projection eigenvalue solver, physically meaningful eigenmodes $\tilde{\mathbf{E}}(\mathbf{r})$ and eigenvalues 
$\tilde{\omega}$ in a specific region in the complex plane can be computed.
Therefore, a contour $\tilde{C}$ around the area of interest has to be chosen. 
Let the electric field $\mathbf{E}(\mathbf{r},\omega_0)$ be a scattering solution to 
Eq.~\eqref{eq:TimeHarmonicMaxwell}.
Let further $\mathcal{L}(\mathbf{E}(\mathbf{r},\omega_0))$ be a physical observable.
Then, the eigenvalues $\tilde{\omega}$ are the resonance poles of the corresponding complex continuation
$\mathcal{L}(\mathbf{E}(\mathbf{r},\omega \in \mathbb{C}))$
into the complex plane. 
Note that physical light sources are typically driven at real frequencies, 
while for the complex continuation we solve scattering problems with sources at complex frequencies.
In the following, for simplicity, a contour is considered which encloses only one eigenvalue $\tilde{\omega}$.
The Laurent expansion of $\mathcal{L}(\mathbf{E}(\mathbf{r},\omega))$ around the eigenvalue $\tilde{\omega}$ with order $p$ is given by
\begin{equation}
    \mathcal{L}(\mathbf{E}(\mathbf{r},\omega)) = \sum_{k=-p}^{\infty} a_k (\omega - \Tilde{\omega})^k \hspace{0.5cm} \text{with} \hspace{0.5cm} a_k(\Tilde{\omega}) = \frac{1}{2 \pi i} \oint\limits_{\Tilde{C}} \frac{\mathcal{L}(\mathbf{E}(\mathbf{r},\omega))}{(\omega - \Tilde{\omega})^{k+1}} \text{d}\omega \hspace{0.5cm} \in \mathbb{C}.
\end{equation}
This expansion, the assumption of $p=1$, and the application of the integral formula of Cauchy lead to
\begin{equation}
    \Tilde{\omega} = \frac{\oint\limits_{\Tilde{C}}\omega \mathcal{L}(\mathbf{E}(\mathbf{r},\omega))\text{d}\omega}{\oint\limits_{\Tilde{C}}\mathcal{L}(\mathbf{E}(\mathbf{r},\omega))\text{d}\omega}.
    \label{eq:Eigenfrequency}
\end{equation}
To numerically compute the eigenvalue $\tilde{\omega}$, the contour in Eq.~\eqref{eq:Eigenfrequency} 
is sampled at discrete $\omega$ values. 
For all of these values, scattering problems are solved, 
and $\mathcal{L}(\mathbf{E}(\mathbf{r},\omega))$ corresponds to a point evaluation of the computed electric field. 
The integration along the contour is realized using the trapezoidal rule. 

In the case of multiple eigenvalues with arbitrary order $p$ within the contour, the formulas are slightly more complex, 
resulting in a nonlinear system of equations. 
The reader is referred to Ref.~\cite{binkowski2020riesz} for a detailed derivation of the Riesz projection eigenvalue solver.
The solver, including the general case of multiple eigenvalues, is implemented in the software package 
RPExpand \cite{betz2021rpexpand}.

\section{Numerical Study}
\label{sec:NumericalAnalysis}
\subsection{VCSEL model}
\label{subsec:ModelProblem}
The VCSEL studied here is shown  as a schematic in Fig.~\ref{fig:FigVCSELSchemea}. Basically, it consists of a substrate, a bottom distributed Bragg reflector (DBR), a cavity, and a top DBR through which the light is emitted to the superspace. 
The diameter of the whole structure is given as $D=30\,\mu\text{m}$. 
The DBR mirrors consist of pairs of Al$_{0.9}$Ga$_{0.1}$As/GaAs layers. 
The linear grading between the two materials is represented in the discrete setup by an intermediate layer with a constant permittivity, averaged over the continuous material profile in the grading area. 
The cavity is limited in the radial direction by the closed part of the oxide aperture. 
The open part of the aperture is limited by the radius $w_{oa} = 3\,\mu\text{m}$. In the center of the cavity, there are five quantum well films with active material (MQWs) surrounded by six buffer layers. At the top and bottom, this area is completed with interface layers up to the oxide aperture. Detailed geometry and material information are given in Tab.~\ref{tab:GeometryAndMaterial}.  
\begin{table}[]
    \centering
    \begin{tabular}{|l|l|l|c|c|c|} \hline
\textbf{Domain}    &  \textbf{Specific layer}      & \textbf{Material}      & \textbf{Number of layers} & \textbf{Layer thickness} & $\varepsilon_r$    \\ \hline
        Free space &  air                          & air                    &  1               &  $100\,\text{nm}$    &  1               \\ \hline
        Top DBR    & low index material            & Al$_{0.9}$Ga$_{0.1}$As & 16               & $61.8\,\text{nm}$    &  9.1494          \\
                   & high index material           & GaAs                   & 15               & $52.6\,\text{nm}$    & 12.127           \\
                   & grading layer                 & Al$_{0.45}$Ga$_{0.55}$As  & 31               & $18\,\text{nm}$   & 10.6382          \\ \hline
        Interface layer &  top                     & Al$_{0.8}$Ga$_{0.2}$As    & 1                & $33.89\,\text{nm}$&  9.4981          \\ 
                        & bottom                   & Al$_{0.8}$Ga$_{0.2}$As    & 1                & $28.25\,\text{nm}$&   9.4981         \\ \hline
        Oxide aperture  & open ($w_{oa}=3\,\mu\text{m}$)&  Al$_{0.98}$Ga$_{0.02}$As & 2                & $20\,\text{nm}$   & 8.8727           \\ 
                        & closed                   & AlOx                      & 2                & $20\,\text{nm}$   & 2.56              \\ \hline 
        Cavity/MQW      &  active layer       & In$_{0.23}$Ga$_{0.77}$As       & 5                & $4\,\text{nm}$    & 15.3343           \\
                        &  barriers           & GaAs$_{0.86}$P$_{0.14}$        & 6                & $5\,\text{nm}$    & 10                \\ \hline
         Bottom DBR     & low index material  & Al$_{0.9}$Ga$_{0.1}$As         & 37               & $61.8\,\text{nm}$ &  9.1494           \\
                        & high index material & GaAs                           & 37               & $52.6\,\text{nm}$ & 12.127            \\
                        & grading layer       & Al$_{0.45}$Ga$_{0.55}$As       & 74               & $18\,\text{nm}$   & 10.6382           \\ \hline
    \end{tabular}

    \caption{Geometry and material data of the VCSEL considered in the numerical study.}
    \label{tab:GeometryAndMaterial}
\end{table}

\begin{figure}
    \centering
    \subfloat[]{\input{Bilder/Arnoldi_Eigenvalues.tex}\label{fig:ArnoldiSolutiona}} \hspace{0.2cm}
    \subfloat[]{\includegraphics[scale=0.135]{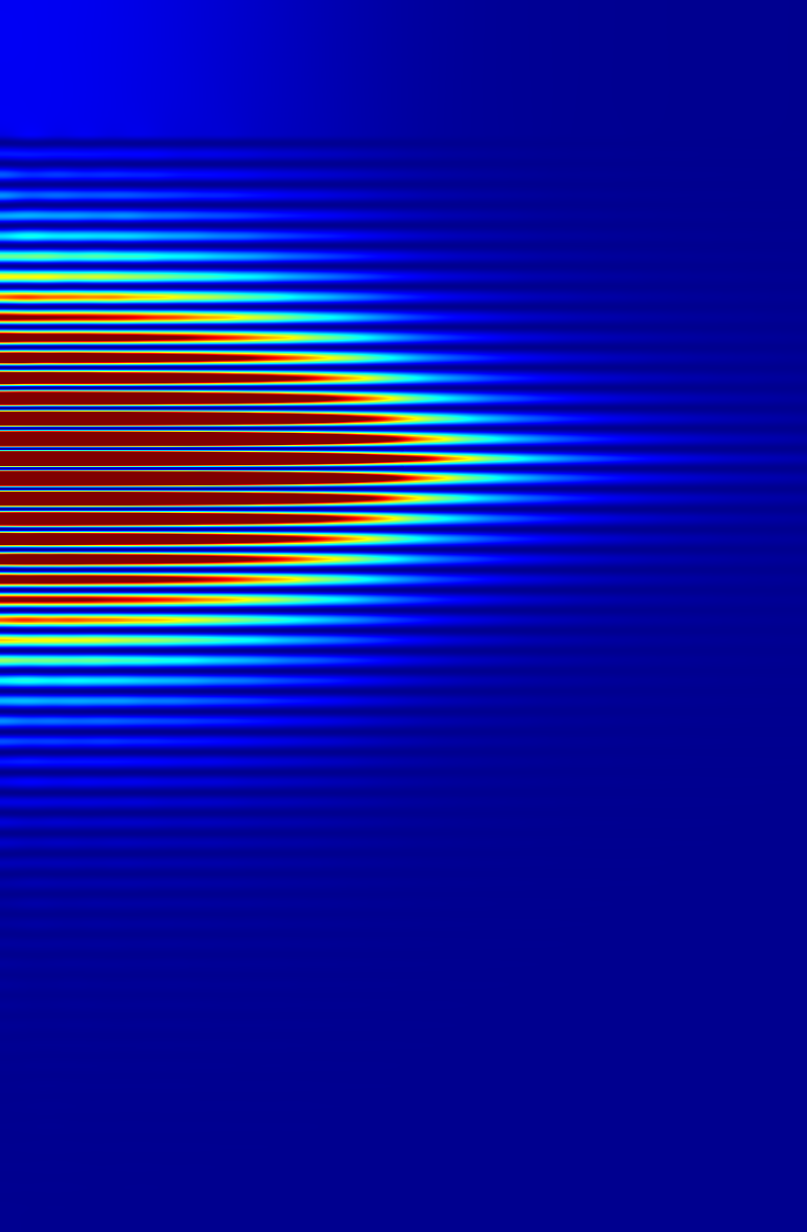}\label{fig:ArnoldiSolutionb}}\hspace{0.1cm}
    \subfloat[]{\includegraphics[scale=0.135]{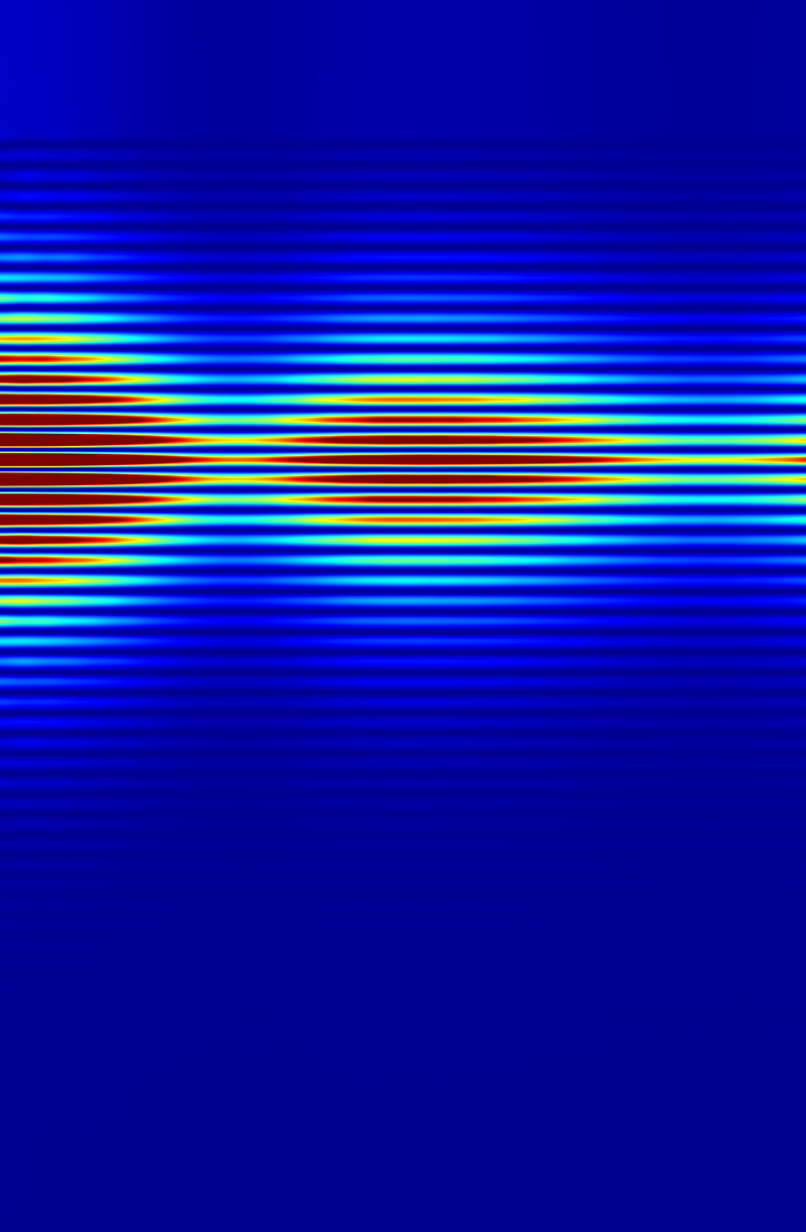}\label{fig:ArnoldiSolutionc}} \hspace{0.1cm}
    \subfloat[]{\includegraphics[scale=0.135]{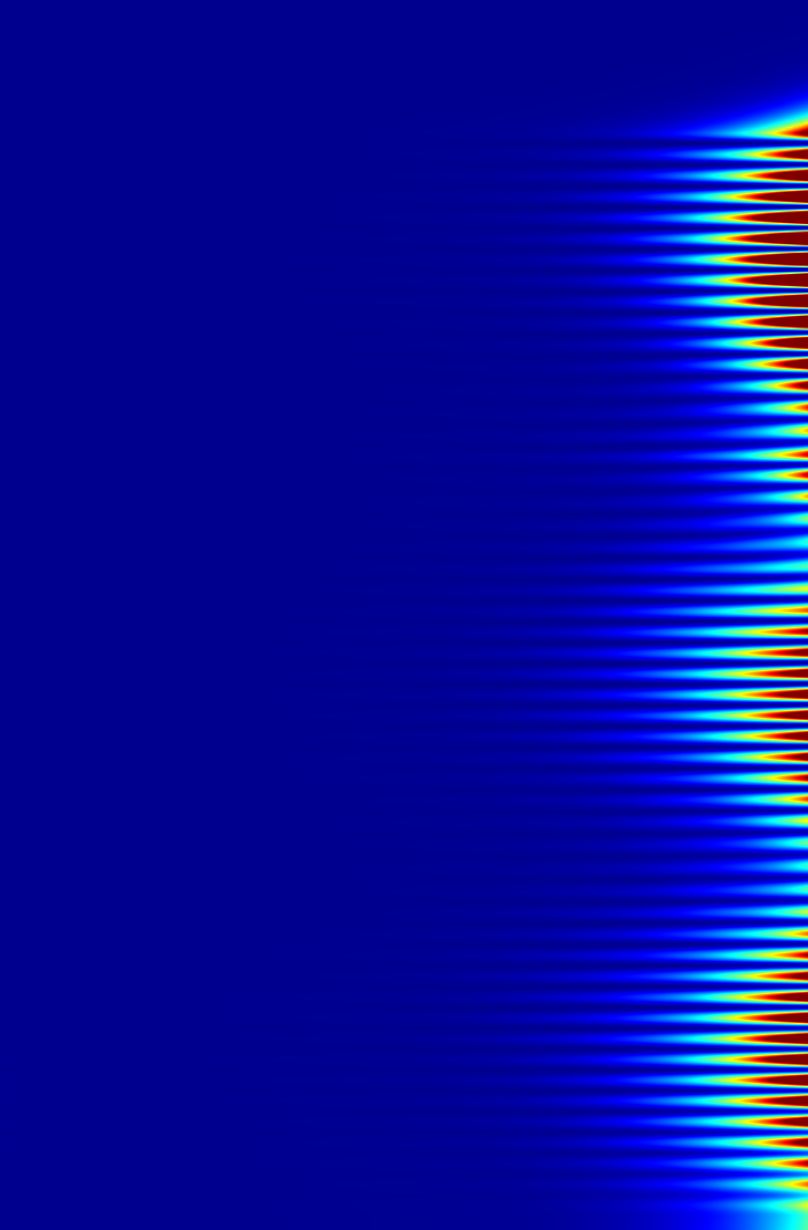}\label{fig:ArnoldiSolutiond}}
    \caption{(a) Complex eigenvalues computed with the Arnoldi method.  
    According to the evaluation of the eigenmodes, the fundamental mode is indicated with a red star, 
    and all the other eigenvalues with a blue circle. 
    (b) Electric field intensity of the fundamental resonance mode. 
    (c) Electric field intensity of the eigenmode next to the fundamental mode. 
    (d) Electric field intensity of an eigenmode that corresponds to a non-physical solution. 
    For all pseudocolor field images in (b) - (d), high field intensities are shown in red and low intensities in blue, 
    at different scales. Here, a cross section in $x$-$y$-coordinates with the $y$-axis as rotational axis is shown, see Fig.~\ref{fig:FigVCSELSchemeb}.
    }
    \label{fig:ArnoldiSolution}
\end{figure}

\subsection{Simulation setup}
\label{subsec:SimulationSetup}
Here, we use a slightly simplified model of the VCSEL described in Section~\ref{subsec:ModelProblem} which neglects the finite size of 
the layers in radial direction with a mesa diameter of $D = 30\,\mu\text{m}$. 
By comparing simulations of the full structure with a diameter of $D = 30\,\mu\text{m}$ to a structure where the layers are modeled 
to extend to infinity, without BCB material, we find just a slight relative shift of the  eigenvalue of the order of $10^{-5}$.
This justifies using a significantly smaller computational domain with absorbing perfectly matched layer (PML) boundaries\cite{berenger1994perfectly} taking into account the layers of different 
materials also in the exterior domain. 
Further, the model assumes cylindrical symmetry which allows to compute the field distributions on a two-dimensional FEM mesh.  
A sketch of the setup is shown in Fig.~\ref{fig:FigVCSELSchemeb}.

The software package RPExpand \cite{betz2021rpexpand} is applied to compute the eigenvalues, where the FEM solver JCMsuite is used to perform the scattering simulations.
For all simulations, we use a polynomial degree of the FEM ansatz functions of $p_{\mathrm{FE}}=4$. 
Three different source types are examined in more detail for the Riesz projection eigenvalue solver, 
which are differently close to the actual physical source. 
The first source is a plane wave with $x$-polarization, which couples into the VCSEL from above. 
The second source is a dipole in the center of the cavity polarized in the $x$-direction for the setup with a single eigenvalue and polarization in the $x$-$y$-$z$-direction for a contour with various eigenvalues enclosed within the contour. The third source is a constant current which is impressed in the area of all five active layers with the same polarization as for the dipole source. The current source comes closest to the physical excitation by the MQWs, due to its allocation.

\subsection{Reference solution}
\label{subsec:ReferenceSolution}
The reference solution is computed with the Arnoldi method-based \cite{saad2011numerical} eigenvalue solver within JCMsuite. 
Computations are performed with $p_{\mathrm{FE}}=4$ and with a guess value of $\omega = 1.922\cdot 10^{15}\,s^{-1}$. 
The resulting reference eigenfrequency is given by $\omega_{\text{ref}}$.
Forty eigenvalues in the complex plane, in the range of the fundamental resonance of $\lambda = 980\,\text{nm}$ 
are plotted in Fig.~\ref{fig:ArnoldiSolutiona}.  
According to the evaluation of the eigenmodes, the fundamental mode is herein indicated with a red star,  and all of the other eigenmodes with blue circles.
The eigenvalues are very closely spaced. 
This complicates the computation of the fundamental mode since many eigenmodes must be calculated to find one mode of interest. 
Typically, this is done by analyzing the field distributions of the desired modes. 
The intensity of the electric field of the fundamental mode, 
of the eigenmode corresponding to the eigenvalue next to the eigenvalue of the fundamental mode, 
and of an eigenmode that corresponds to a non-physical solution are visualized 
in Fig.~\ref{fig:ArnoldiSolutionb}-\ref{fig:ArnoldiSolutiond}. 
The fundamental mode has its maximum inside the cavity and the intensity decreases in the area of the DBRs. 
A fraction of the light is emitted via the upper mirror, as would be expected here. 
Higher-order modes have increasing numbers of maxima in the radial direction, 
which also shift further into the center of the cavity as the order number increases. 
In the case of the non-physical solutions, 
field maxima may be located away from the cavity region, for example, close to the PML in the $x$-direction.

\begin{figure}
\centering
\subfloat[]{\input{Bilder/1EW_Konvergenz_radus_schema.tex}}
    \subfloat[]{\input{Bilder/1EV_Konvergenzen_Integrationspunkte.tex}\label{fig:EinEVKonvergenza}}\\
    \hspace{2.5cm}     \subfloat[]{\includegraphics[scale=0.15]{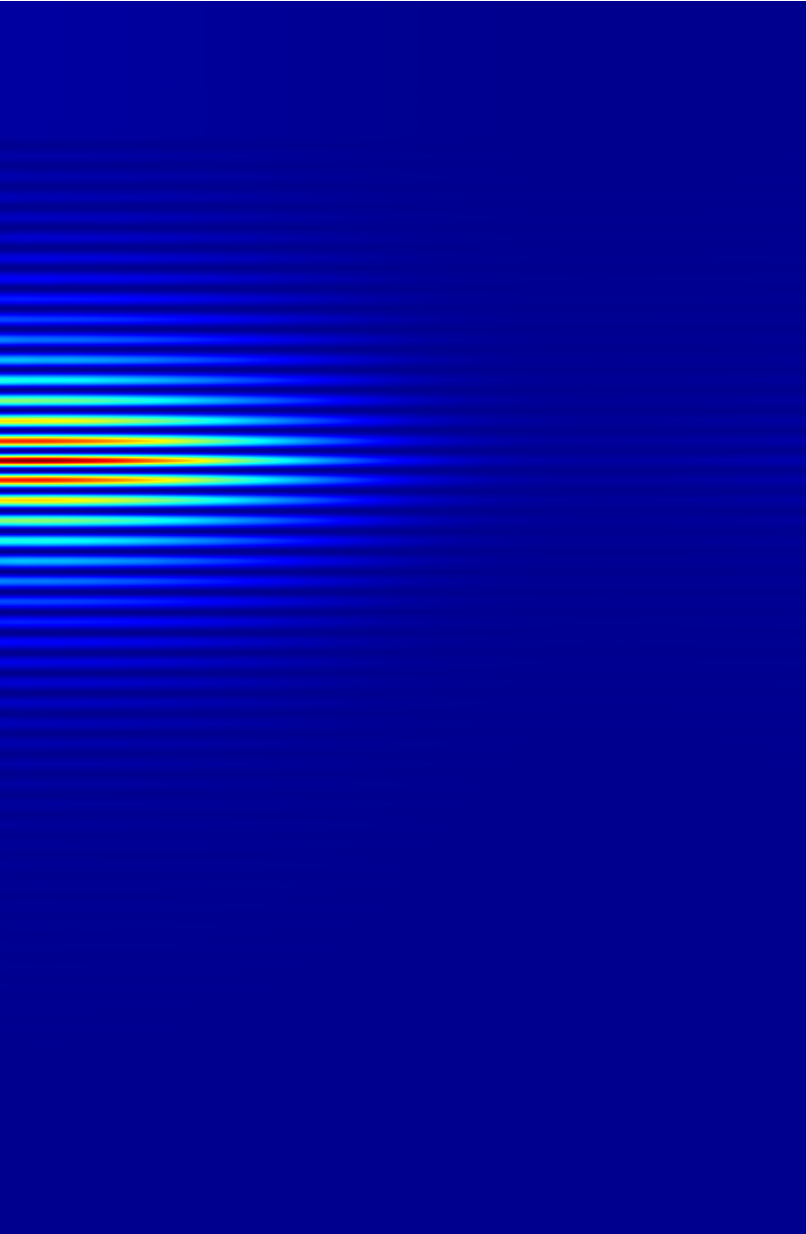}}\hspace{1.2cm}
     \subfloat[]{\input{Bilder/1EV_Konvergenzen_Radius_Kontur_4Punkte.tex}\label{fig:EinEVKonvergenzb}}
    \caption{Computation of the fundamental mode by using the Riesz projection eigenvalue solver where the contour does not contain any further eigenvalues. (a) Integration contour with eigenvalue of the fundamental mode inside and further eigenvalues outside of the contour. With radius of the contour $R$ and minimal distance to the next nearest eigenvalue $d_{\text{min}}$. 
    (b) Convergence of the relative error of the eigenvalue of the fundamental mode with the Riesz projection eigenvalue solver with respect to the number of integration points $N_P$ at a radius of the integration contour of $R = \tfrac{1}{2}  d_{\text{min}}$. 
    (c) Field intensity of the electric field for a converged simulation. The excitation source is the current source. High field intensities are shown in red and low in blue. Here, a cross section in $x$-$y$-coordinates with the $y$-axis as rotational axis is shown, see Fig.~\ref{fig:FigVCSELSchemeb} for details.  
    (d) Convergence of the relative error of the eigenvalue of the fundamental mode with the Riesz projection eigenvalue solver with respect to the radius of the integration contour normalized to the minimal distance to the next nearest eigenvalue $R = [0.01,0.9] \cdot d_{\text{min}}$ for four integration points.}
    \label{fig:EinEVKonvergenz}
\end{figure}

\subsection{Results -- single eigenvalue within the integration contour}
\label{subsec:ResultsSingleEigenvalueWithinIntegrationContour}
Firstly, the Riesz projection eigenvalue solver is tested using a contour in which 
exclusively the fundamental mode is enclosed. 
For this purpose, the radius of the contour is selected as a function of the distance to the next nearest eigenvalue, referred to as $d_{\text{min}}$ in the following. This is obtained from the reference solution, which has already been computed with the Arnoldi method. Three source types, as discussed in Section~\ref{subsec:SimulationSetup}, are investigated in more detail: plane wave, dipole, and constant current.

First, the radius of the contour is fixed at $R = \tfrac{1}{2}  d_{\text{min}}$ 
and the eigenvalues are computed, using different numbers of integration points $N_P$ on the contour. 
For each of these, the relative error of the real and imaginary part of the eigenvalue is computed as 
$\Delta \Re(\omega) = \left| \Re(\Tilde{\omega}) - \Re(\omega_{\text{ref}}) \right| /  \left| \Re(\omega_\text{ref}) \right|$ 
and
$\Delta \Im(\omega) = \left| \Im(\Tilde{\omega}) - \Im(\omega_{\text{ref}}) \right| /  \left| \Im(\omega_\text{ref}) \right|$, 
with the solution of the Riesz projection eigenvalue solver $\Tilde{\omega}$ 
and the reference solution from the Arnoldi method $\omega_{\text{ref}}$. 
The results for all three excitations and for the real and imaginary parts of the eigenvalue  are shown 
in Fig.~\ref{fig:EinEVKonvergenza}. 
Very low relative errors for the real and imaginary parts of the eigenvalue are reached already with relatively low numbers of 
integration points. 
The relative error of the imaginary part is in general larger than that of the real part, due to the  
difference in the size of the real and imaginary part of $10^4$ of the eigenvalue itself. 
The convergence curves for the excitation with a dipole and with constant current are almost identical, 
the relative error decreases fast. 
The convergence curve for the excitation with the plane wave decreases slower. 

Next, the convergence with respect to the radius of the integration contour is examined. 
For this purpose, the number of integration points is fixed to $N_P=4$ and the radius of the contour 
is varied in the range of $[0.01,0.9] \cdot d_{\text{min}}$, respectively. 
The results are shown in Fig.~\ref{fig:EinEVKonvergenzb}. 
For all three setups, the relative error is decreasing with the radius of the integration contour. 
For $0.01 \cdot d_{\text{min}}$, the relative error of the real part of the eigenvalue is 
decreased to $10^{-12}$ and the imaginary part to $10^{-9}$. 
The curves for the three different excitation sources are looking similar. 
This corresponds to the result from the previous convergence study, 
where it can be seen that the relative error for four integration points is similar for all the setups.  

In summary, we can conclude that the numerical error of the computed eigenvalue converges with an 
increasing number of integration points as well 
with decreasing radius of the integration contour.

\begin{figure}
 \subfloat[]{\input{Bilder/4EV_Integrationskontur.tex} \label{fig:4Eigenwertea}}
 \subfloat[]{\input{Bilder/4EV_Konvergenzen_Integrationspunke.tex} \label{fig:4Eigenwerteb}}\\
 \subfloat[]{\input{Bilder/Konvergenz_4EV_N_Expected_Eigenvalues}\label{fig:4Eigenwertec}}
       \caption{(a) Eigenvalues in the complex plane, integration contour in red contains four eigenvalues, Arnoldi solution marked by cyan stars, solution of the Riesz projection eigenvalue solver inside the contour as blue dots. 
       The electric field intensities of the four modes, computed with the Riesz projection eigenvalue solver, 
       are displayed in the bottom of the subfigure. 
       A cross section in $x$-$y$-coordinates with the $y$-axis as rotational axis is shown, see Fig.~\ref{fig:FigVCSELSchemeb}. 
       High field intensities are shown in red and low intensities in blue.
       (b) Convergence of the relative error of the real and imaginary part of the four eigenvalues calculated with the Riesz projection eigenvalue solver depending on the number of integration points $N_P$.
       (c) Convergence of the relative error of the real and imaginary part of the four eigenvalues calculated with the Riesz projection eigenvalue solver depending on the number of unknowns $m$ for the NLSE.
       } 
       \label{fig:4Eigenwerte}
\end{figure}
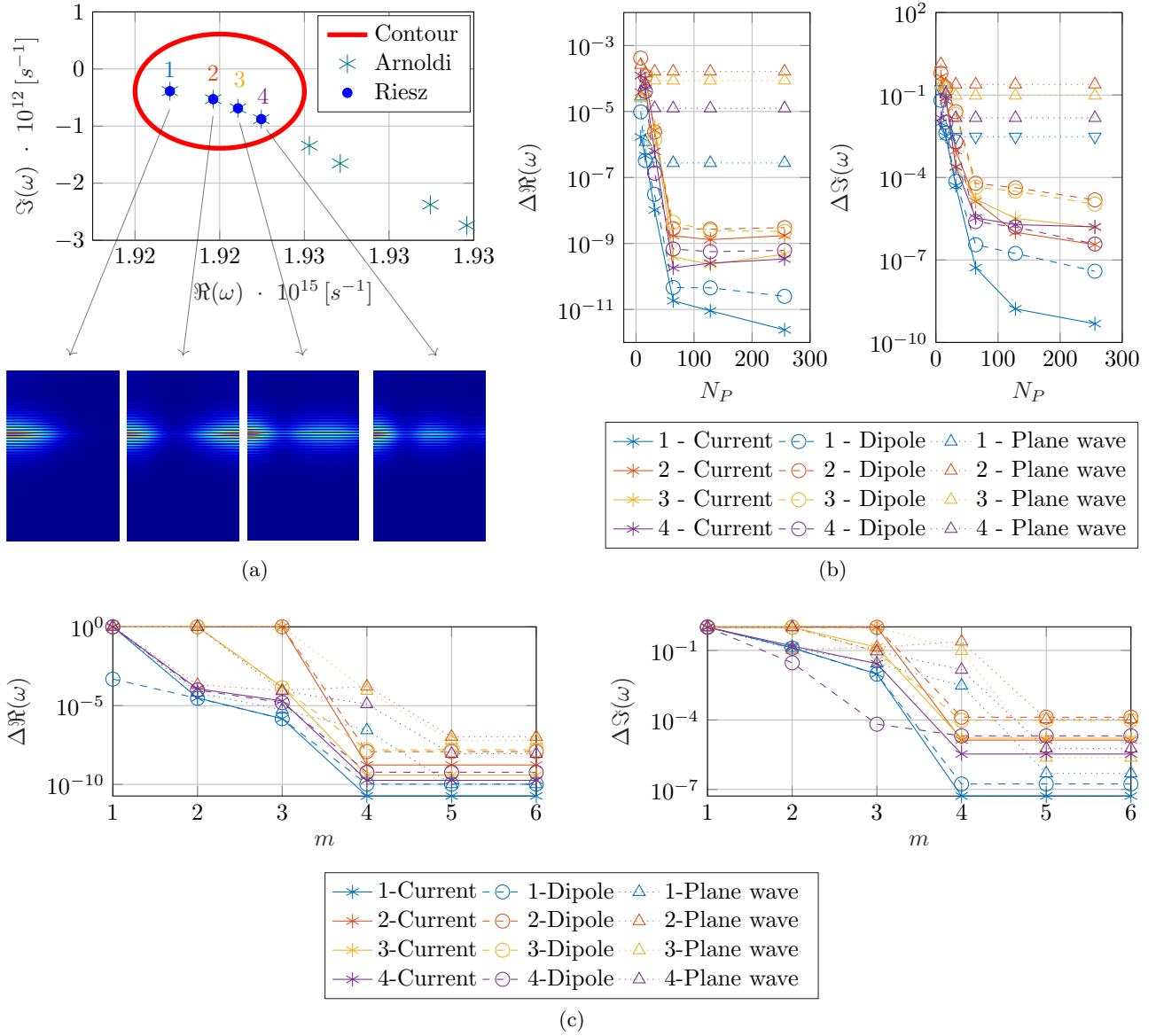

\subsection{Results -- multiple eigenvalues within the integration contour}
\label{subsec:ResultsMultipleEigenvaluesWithinIntegrationContour}
For this study, the integration contour is chosen such that it encloses the fundamental mode and three other eigenvalues, 
see Fig.~\ref{fig:4Eigenwertea}. 
The Figure shows the integration contour in the complex plane with the reference values 
of the eigenvalues and the converged solutions of the Riesz projection eigensolver 
with excitation via a current within the active material layers.  
In addition, the intensity of the electric field strength for all four eigenvalues is shown in Fig.~\ref{fig:4Eigenwertea}. 
It can be seen that the converged values for the Riesz projection eigenvalue solver are in excellent agreement 
with the reference values obtained by the Arnoldi method. 
Furthermore, the field intensities of all four modes are in good agreement with the reference solutions.
The convergence of the different modes with respect to the number of integration points on the contour is investigated and 
shown in Fig.~\ref{fig:4Eigenwerteb}.  

For the excitation with a current inside the active layers and the excitation with a dipole inside the cavity, 
a very similar convergence behavior results for all four modes. 
While in some cases eight integration points are sufficient to find the four eigenvalues within the contour, 
in others at least 16 points are required. 
Very low relative errors of about $10^{-9}$ and below are achieved already for 64 integration points. 
For both excitations, for real and imaginary parts, the eigenvalue of the fundamental mode 
achieves smaller errors and has better convergence behavior than the higher order modes. 
For the excitation with a plane wave, the error convergence is not as good. 
Fig.~\ref{fig:4Eigenwertec} shows the convergence of the relative error for the real and imaginary parts of the eigenfrequency, depending on $m$, the number of unknowns for the nonlinear system of equations (NLSE). 
This limits the size of the nonlinear system of equations to be solved. 
For calculating a specific number of eigenvalues within the contour the number of expected eigenvalues should be equal or larger than the number of enclosed eigenvalues.
However, when this number is not known, the $m$ can in principle be adapted automatically. 

The first eigenfrequency which is found for all three excitations is that of the fundamental mode, it has the best convergence behavior. 
As before, the convergence behavior of current excitation and dipole source is almost identical. 
For the use of the plane wave, one obtains larger relative errors.  
The fundamental mode can be determined 
by its convergence behavior or by the value of the corresponding residue \cite{binkowski2020riesz}. 
The residue of the fundamental mode is largest with respect to the other eigenvalues within the contour. 
This is because the excitations couple best to the fundamental mode. 

Overall, it can be stated that with the appropriate choice of the excitation source 
for the scattering simulation, very good results are obtained with the Riesz projection eigenvalue solver. 
This holds for a single eigenvalue as well as for several eigenvalues within the contour. 
The convergence of the eigenvalues depends on the number of integration points 
along the contour, on the contour size, and on the excitation source chosen for the scattering problem.

\section{Conclusions}
\label{sec:Conclusions}
We have analyzed the computation of quasi-normal modes and corresponding 
eigenvalues of a VCSEL using a Riesz projection eigenvalue solver.  
First, an integration contour was chosen in which exclusively the eigenvalue of the fundamental mode exists.  
For this configuration, the convergence of the eigenvalue with respect to the number of integration points 
for a fixed radius and the radius of the contour with a fixed number of integration points were investigated. 
Three different sources were used as excitation for the scattering simulations: a current inside the active material layers, a dipole in the center of the cavity, and a plane wave coupled from above the VCSEL. It was shown that the eigenvalue converges with an increasing number of integration points at a fixed radius of the integration contour. The eigenvalue calculation is also converging for a contour with increasing distance to the next eigenvalue outside of the contour. 
No significant differences between the results obtained with the three different excitations have been observed.

Further, a contour has been examined in which four eigenvalues are located. 
The convergence of all these eigenvalues with respect to the number of integration points for a fixed contour was analyzed as well.
Convergence could be shown for all four eigenvalues. 
The results for the excitation with a constant current and with a dipole were similar. 
For the excitation with a plane wave, significantly larger errors were obtained. 
The smallest errors occur for the fundamental mode. 
This can be explained by the fact that all sources chosen in this study couple best to this mode. 
Also, as expected, the field patterns of the modes,  obtained with the Riesz projection eigenvalue solver 
correspond to the reference simulation results.

Eigenvalue computations with the Arnoldi method require linearization in the case of dispersive materials, resulting in high memory consumption. 
With the Riesz-projection-based eigenvalue solver, on the other hand, multiple scattering problems have to be computed. These are smaller in terms of memory and can be solved in parallel. 
With the Arnoldi method, it is straightforward to compute a larger number of eigenvalues. 
With the Riesz projection method usually, only a few eigenvalues are computed. 
However, it is possible to calculate solely eigenvalues with physical meaning\cite{binkowski2020riesz}. 
Furthermore, partial derivatives, resp.~sensitivities of the eigenvalues \cite{binkowski2022computation} can be computed 
with negligible additional effort. 
These can be used, e.g., to improve the convergence of optimization problems.
Also, modal expansions of further derived quantities\cite{betz2021rpexpand, zschiedrich2018riesz}, like scattering cross sections or
far-field fluxes, can be performed in this framework. 
As the approach relies on scattering simulations, the Riesz projection eigenvalue solver allows us to 
solve eigenvalue problems on domains that can be decomposed into subdomains of different symmetries, 
by applying domain decomposition methods \cite{Schaedle_jcp_2007}.
This may, e.g., enable to efficiently solve the eigenvalue problem for VCSELs where the top mirror is 
realized with a 1D or 2D diffraction grating\cite{Gebski:19}.  

\label{sec:misc}
\acknowledgments 
This project is co-financed by the European Regional Development Fund (EFRD, application no. 10184206, QD-Sense).
This project (20FUN05 SEQUME) has received funding from the EMPIR programme co-financed by the Participating States and from the European Union’s Horizon 2020 research and innovation programme.
 Funded by the Deutsche Forschungsgemeinschaft (DFG, German Research Foundation) under Germany's Excellence Strategy – The Berlin Mathematics Research Center MATH+ (EXC-2046/1, project ID: 390685689).
 This project has received funding from the German Federal Ministry of Education and Research (BMBF Forschungscampus  MODAL, project number 05M20ZBM).

% References
%\bibliography{report} % bibliography data in report.bib
%\bibliographystyle{spiebib} % makes bibtex use spiebib.bst

\end{document}

%% file: Bilder/SchematischeDarstellung.tex
\begin{tikzpicture}[ axis/.style={very thick, ->, >=stealth'}, pile/.style={thick, ->, >=stealth', shorten <=2pt, shorten >=2pt}, scale=1]
   \draw[transparent] (0,-0.5) rectangle(3,-0.9);		
   \draw[fill=blue!10!white, draw =blue!10!white] (-3.5,-0.02) rectangle (3.5, 4.26);
       
   \draw[fill=gray!70!white, draw = gray!70!white] (-3.5,-0.5) rectangle (3.5,-0.03);
   \draw (0, -0.25) node{GaAs Buffer};
	
    %\draw[fill=gray!10!white] (-4,0) rectangle (4,3);
    \foreach \y in {0,0.2,0.4, 0.6, 0.8, 1, 1.2, 1.4, 1.6, 1.8}
    \draw[fill=gray!10!white, draw=gray!10!white] (-2.5,\y) rectangle (2.5,\y+0.1);

    \foreach \y in {0.1,0.3,0.5, 0.7, 0.9, 1.1, 1.3, 1.5, 1.7, 1.9}
    \draw[fill=gray!50!white, draw=gray!50!white] (-2.5,\y) rectangle (2.5,\y+0.1);

    %\draw[fill=gray!10!white] (-3,3) rectangle (3,5);
    \foreach \y in {2.02,2.22,2.42,2.62}
    \draw[fill=gray!10!white, draw=gray!10!white] (-2.5,\y) rectangle (2.5,\y+0.1);

    \foreach \y in {2.12,2.32,2.52,2.72}
    \draw[fill=gray!50!white, draw=gray!50!white] (-2.5,\y) rectangle (2.5,\y+0.1);
    \draw (0,1.35) node{Bottom DBR};    
    
    % Cavity
    \draw[fill=gray!10!white, draw=gray!10!white] (-2.5,2.84) rectangle (2.5, 3.36);

    % Top DBR
    \foreach \y in {3.36,3.56,3.76,3.96,4.16}
    \draw[fill=gray!50!white, draw=gray!50!white] (-2.5,\y) rectangle (2.5,\y+0.1);
    
    \foreach \y in {3.46,3.66,3.86,4.06}
    \draw[fill=gray!10!white, draw=gray!10!white] (-2.5,\y) rectangle (2.5,\y+0.1);
    \draw (0,3.9) node{Top DBR};  
    
   % Oxid Apertur
   \draw[fill=black] (-2.5,2.84) rectangle (-1, 2.94);
   \draw[fill=black] (2.5,2.84) rectangle (1, 2.94);
   \draw[fill=black] (-2.5,3.24) rectangle (-1, 3.34);
   \draw[fill=black] (2.5,3.24) rectangle (1, 3.34);

   % active layers
   \draw[black] (-2.5, 3.04) -- (2.5, 3.04);
   \draw[black] (-2.5, 3.09) -- (2.5, 3.09);
   \draw[black] (-2.5, 3.14) -- (2.5, 3.14);

   % Nodes for MQW, Oxide Aperture, BCB and air
   \draw[<-] (2.6,3.09) -- (3.75,3.09) node[above, xshift=-3ex]{MQW};
   
   \draw[->] (-3.25,2.91) -- (-2.65,2.91);
   \draw[->] (-3.25, 3) -- (-2.65,3.25);
   \draw(-4.5,3.35) node{Oxide};
   \draw(-4.3,2.95) node{Aperture};

   \draw (-3,0.25) node{BCB};
   \draw ( 3,0.25) node{BCB};
   \draw (-3,4.5) node{Air};

   \draw[<->] (-2.5,0.25) -- (2.5,0.25) node[above,xshift=-16ex]{$D=30\,\mu \text{m}$};
   \draw[<->] (-1,2.7) -- (1,2.7) node[below,xshift=-6ex]{$2w_{oa}=6\,\mu\text{m}$};
\end{tikzpicture}

%% file: Bilder/SchematischeDarstellung_NumerischesSetup.tex
\begin{tikzpicture}[ axis/.style={very thick, ->, >=stealth'}, pile/.style={thick, ->, >=stealth', shorten <=2pt, shorten >=2pt}, scale=1]
   		
   \draw[fill=gray!70!white, draw = gray!70!white] (0,-0.5) rectangle (3.5,-0.03);
   
	% Bottom DBR
    \foreach \y in {0,0.2,0.4, 0.6, 0.8, 1, 1.2, 1.4, 1.6, 1.8}
    \draw[fill=gray!10!white, draw=gray!10!white] (0,\y) rectangle (3.5,\y+0.1);

    \foreach \y in {0.1,0.3,0.5, 0.7, 0.9, 1.1, 1.3, 1.5, 1.7, 1.9}
    \draw[fill=gray!50!white, draw=gray!50!white] (0,\y) rectangle (3.5,\y+0.1);
    
    \foreach \y in {2.02,2.22,2.42,2.62}
    \draw[fill=gray!10!white, draw=gray!10!white] (0,\y) rectangle (3.5,\y+0.1);

    \foreach \y in {2.12,2.32,2.52,2.72}
    \draw[fill=gray!50!white, draw=gray!50!white] (0,\y) rectangle (3.5,\y+0.1);

    % Cavity
    \draw[fill=gray!10!white, draw=gray!10!white] (0,2.84) rectangle (3.5, 3.36);

    % Top DBR
    \foreach \y in {3.36,3.56,3.76,3.96,4.16}
    \draw[fill=gray!50!white, draw=gray!50!white] (0,\y) rectangle (3.5,\y+0.1);
    
    \foreach \y in {3.46,3.66,3.86,4.06}
    \draw[fill=gray!10!white, draw=gray!10!white] (0,\y) rectangle (3.5,\y+0.1);

     % Oxid Apertur
   \draw[fill=black] (3.5,2.84) rectangle (1.75, 2.94);
   \draw[fill=black] (3.5,3.24) rectangle (1.75, 3.34);

   % active layers
   \draw[black] (0, 3.04) -- (3.5, 3.04);
   \draw[black] (0, 3.09) -- (3.5, 3.09);
   \draw[black] (0, 3.14) -- (3.5, 3.14);

   \draw[->, thick] (0,-.5) -- (4,-0.5) node[below, xshift=-2ex]{$x$};
   \draw[->, thick] (0,-0.5) -- (0,5.5) node[left, yshift=-3ex]{$y$};

   \draw[fill=blue, semitransparent] (0,-0.5) -- (3.5,-0.5) -- (3.5,5) -- (0,5) -- (0,4.8) -- (3.3,4.8) -- (3.3,-0.3) -- (0,-0.3) -- (0,-0.5);
   \draw(4.3, 1.2) node{PML};
   \draw[<-] (3.55,1.2) -- (3.9,1.2); 
   % Sources 1
     \draw[yellow, fill=yellow, -{Triangle[width = 18pt, length = 8pt]}, line width = 6pt] (1, 5.55) -- (1.0, 5.05) node[right, black, xshift=2ex, yshift=2ex]{1: Plane wave};
    
    % Source 3
    \draw[red, thick] (0, 3.04) -- (3.5, 3.04);
    \draw[red, thick] (0, 3.09) -- (3.5, 3.09);
    \draw[red, thick] (0, 3.14) -- (3.5, 3.14);
    \draw (1.4,1.8) node{3: Current source};
    \draw[->] (0.75, 2.1) --(0.75, 3);

    % Sorce 2
    \draw[yellow, -Triangle, very thick] (0, 3.09) -- (0.3, 3.09);
    \draw[yellow, -Triangle, very thick] (0, 3.09) -- (-0.3, 3.09);
    \draw (1.3,3.8) node{2: Dipole source};
    \draw[->] (0.75,3.5) -- (0.1,3.25);

    \draw[<->] (0,0.25) -- (3.3,0.25) node[above, xshift=-11ex]{$D_{xy}/2 = 6\,\mu\textbf{m}$};
\end{tikzpicture}

%% file: Bilder/Arnoldi_Eigenvalues.tex
% This file was created by matlab2tikz.
%
%The latest updates can be retrieved from
%  http://www.mathworks.com/matlabcentral/fileexchange/22022-matlab2tikz-matlab2tikz
%where you can also make suggestions and rate matlab2tikz.
%
\definecolor{mycolor1}{rgb}{0.00000,0.44700,0.74100}%
\definecolor{mycolor2}{rgb}{0.85000,0.32500,0.09800}%
\definecolor{mycolor3}{rgb}{0.92900,0.69400,0.12500}%
\begin{tikzpicture}

\begin{axis}[%
width=2.3in,
height=1.05in,
at={(0.772in,0.48in)},
scale only axis,
xmin=1.88,
xmax=1.97,
xlabel style={font=\color{white!15!black}},
xlabel={$\Re\text{(}\omega\text{) }\cdot\text{ 10}^{\text{15}}\, [s^{-1}]$},
ymin=-45,
ymax=0,
ylabel style={font=\color{white!15!black}},
ylabel={$\Im\text{(}\omega\text{) }\cdot\text{ 10}^{\text{12}}\, [s^{-1}]$},
axis background/.style={fill=white},
xmajorgrids,
ymajorgrids,
legend columns=2,
legend style={at={(0,-0.65)}, anchor=south west, legend cell align=left, align=left, draw=white!15!black}
] %at={0.4,0.05}
\addplot [color=mycolor2, only marks, mark size=3.5pt, mark=asterisk, mark options={solid, mycolor2}]
  table[row sep=crcr]{%
1.92282292056358	-0.388784029895019\\
};
\addlegendentry{Fundamental mode   }

\addplot [color=mycolor1, only marks, mark size=2.5pt, mark=o, mark options={solid, mycolor1}]
  table[row sep=crcr]{%
1.92384638620298	-0.52822736905899\\
1.92443133648057	-0.688913018274051\\
1.92498223942906	-0.878120032959705\\
1.92611689426939	-1.34318523911849\\
1.92684635316196	-1.65436077590087\\
1.92897611973415	-2.3763252276247\\
1.92983644339416	-2.73852685433047\\
1.93252500756729	-3.53996001161088\\
1.93335133639161	-3.7256570529773\\
1.93722355010064	-5.23545718593272\\
1.93748645281924	-5.52128336606568\\
1.94222666803619	-6.61114679757787\\
1.94299842582991	-7.12892620366516\\
1.94816642375911	-8.87947506087725\\
1.94859573987747	-9.78463219632112\\
1.95512034015103	-11.8851829560223\\
1.95602613894664	-9.95459168606568\\
1.96185998078294	-15.446188307289\\
1.96256719730153	-13.0269652852053\\
1.88435358954907	-18.6904964040858\\
1.88683918291983	-19.562741356577\\
1.89406289496415	-21.7199715070204\\
1.89687122677021	-23.4162063556948\\
1.89957950646299	-23.3262658160721\\
1.90116221841542	-26.0212289330424\\
1.90518728598256	-38.6082639715345\\
1.90757311567895	-25.887255178748\\
1.90787915043493	-41.2190530320249\\
1.9091759219846	-33.0537028232347\\
1.91278645606024	-28.3966858037575\\
1.91700769187401	-40.5364041288425\\
1.91771918439687	-31.1765463516003\\
1.91880373391378	-44.0209204308347\\
1.92466551256318	-31.6959725962277\\
1.9288728140592	-28.1309560841525\\
1.93092204724639	-35.4001989079887\\
1.93434597273324	-38.6954075566184\\
1.93977951440436	-35.4039018572635\\
1.94102637027389	-37.8795765478245\\
};
\addlegendentry{Eigenvalues }

%\addplot [color=mycolor3, only marks, mark size=2.5pt, mark=triangle, mark options={solid, mycolor3}]
%  table[row sep=crcr]{%
%1.88435358954907	-18.6904964040858\\
%1.88683918291983	-19.562741356577\\
%1.89406289496415	-21.7199715070204\\
%1.89687122677021	-23.4162063556948\\
%1.89957950646299	-23.3262658160721\\
%1.90116221841542	-26.0212289330424\\
%1.90518728598256	-38.6082639715345\\
%1.90757311567895	-25.887255178748\\
%1.90787915043493	-41.2190530320249\\
%1.9091759219846	-33.0537028232347\\
%1.91278645606024	-28.3966858037575\\
%1.91700769187401	-40.5364041288425\\
%1.91771918439687	-31.1765463516003\\
%1.91880373391378	-44.0209204308347\\
%1.92466551256318	-31.6959725962277\\
%1.9288728140592	-28.1309560841525\\
%1.93092204724639	-35.4001989079887\\
%1.93434597273324	-38.6954075566184\\
%1.93977951440436	-35.4039018572635\\
%1.94102637027389	-37.8795765478245\\
%};
%\addlegendentry{Non physical modes}
\addplot [color=mycolor2, only marks, mark size=3.5pt, mark=asterisk, mark options={solid, mycolor2}]
  table[row sep=crcr]{%
1.92282292056358	-0.388784029895019\\
};
\end{axis}
\draw[->] (4,2.9) -- (4.6,3.7);
\draw (4,2.65) node{(b)};

\draw[->] (4.8,2.9) -- (4.9,3.6);
\draw (4.8,2.65) node{(c)};

\draw[->] (5.6,2.4) -- (4.52,1.52);
\draw (5.6,2.65) node{(d)};
\end{tikzpicture}%

%% file: Bilder/1EW_Konvergenz_radus_schema.tex
% This file was created by matlab2tikz.
%
%The latest updates can be retrieved from
%  http://www.mathworks.com/matlabcentral/fileexchange/22022-matlab2tikz-matlab2tikz
%where you can also make suggestions and rate matlab2tikz.
%
\definecolor{mycolor1}{rgb}{0.00000,0.44700,0.74100}%
\definecolor{mycolor2}{rgb}{0.85000,0.32500,0.09800}%
\definecolor{mycolor3}{rgb}{0.92900,0.69400,0.12500}%
\begin{tikzpicture}

\begin{axis}[%
width=2in,
height=1.5in,
at={(0.78in,0.478in)},
scale only axis,
xmin=1.922,
xmax=1.925,
xlabel style={font=\color{white!15!black}},
xlabel={$\Re(\omega) \cdot 10^{15}\, [s^{-1}]$},
ymin=-1.2,
ymax=0.9,
ylabel style={font=\color{white!15!black}},
ylabel={$\Im(\omega) \cdot 10^{12}\, [s^{-1}]$},
axis background/.style={fill=white},
xmajorgrids,
ymajorgrids,
legend style={legend cell align=left, align=left, draw=white!15!black}
]
\addplot [color=red, line width = 2pt]
  table[row sep=crcr]{%
1.92333938120096	-0.388784029894993\\
1.92333689430183	-0.338162035125897\\
1.92332945755463	-0.288027557839224\\
1.92331714257931	-0.238863420457539\\
1.9233000679758	-0.191143100499679\\
1.92327839818189	-0.145326170733417\\
1.92325234188951	-0.101853873238485\\
1.92322215003503	-0.0611448700041335\\
1.92318811338249	-0.0235912109853148\\
1.92315055972347	0.010445441552702\\
1.92310985072023	0.0406372960396488\\
1.92306637842274	0.0666935884122695\\
1.92302056149297	0.0883633823311979\\
1.92297284117302	0.10543798583508\\
1.92292367703563	0.117752961158246\\
1.92287354255835	0.125189708356284\\
1.92282292056358	0.127676607488336\\
1.92277229856881	0.125189708356284\\
1.92272216409152	0.117752961158246\\
1.92267299995414	0.10543798583508\\
1.92262527963418	0.0883633823311979\\
1.92257946270442	0.0666935884122696\\
1.92253599040692	0.0406372960396488\\
1.92249528140369	0.0104454415527021\\
1.92245772774467	-0.0235912109853147\\
1.92242369109213	-0.0611448700041335\\
1.92239349923764	-0.101853873238485\\
1.92236744294527	-0.145326170733417\\
1.92234577315135	-0.191143100499679\\
1.92232869854785	-0.238863420457539\\
1.92231638357252	-0.288027557839224\\
1.92230894682533	-0.338162035125897\\
1.92230645992619	-0.388784029894993\\
1.92230894682533	-0.439406024664089\\
1.92231638357252	-0.489540501950761\\
1.92232869854785	-0.538704639332446\\
1.92234577315135	-0.586424959290307\\
1.92236744294527	-0.632241889056569\\
1.92239349923764	-0.675714186551501\\
1.92242369109213	-0.716423189785852\\
1.92245772774467	-0.753976848804671\\
1.92249528140369	-0.788013501342688\\
1.92253599040692	-0.818205355829634\\
1.92257946270442	-0.844261648202255\\
1.92262527963418	-0.865931442121184\\
1.92267299995414	-0.883006045625066\\
1.92272216409152	-0.895321020948232\\
1.92277229856881	-0.902757768146269\\
1.92282292056358	-0.905244667278322\\
1.92287354255835	-0.902757768146269\\
1.92292367703563	-0.895321020948232\\
1.92297284117302	-0.883006045625066\\
1.92302056149297	-0.865931442121184\\
1.92306637842274	-0.844261648202255\\
1.92310985072023	-0.818205355829635\\
1.92315055972347	-0.788013501342688\\
1.92318811338249	-0.753976848804671\\
1.92322215003503	-0.716423189785852\\
1.92325234188951	-0.675714186551501\\
1.92327839818189	-0.632241889056569\\
1.9233000679758	-0.586424959290307\\
1.92331714257931	-0.538704639332446\\
1.92332945755463	-0.489540501950762\\
1.92333689430183	-0.439406024664089\\
1.92333938120096	-0.388784029894993\\
};
\addlegendentry{Integration Contour}
\addplot [color=mycolor3, only marks,mark size = 5pt, mark=asterisk, mark options={solid, mycolor3}]
  table[row sep=crcr]{%
1.92282292056358	-0.388784029894993\\
};
\addlegendentry{Fundamental mode}

\addplot [color=mycolor2, only marks, mark size = 5pt, mark=asterisk, mark options={solid, teal}]
  table[row sep=crcr]{%
1.92384638620298	-0.528227369059039\\
1.92443133648057	-0.688913018274724\\
};
\addlegendentry{Eigenvalues}
\addplot [color=mycolor3, only marks,mark size = 5pt, mark=asterisk, mark options={solid, mycolor3}]
  table[row sep=crcr]{%
1.92282292056358	-0.388784029894993\\
};
\end{axis}
\draw[->] (3.4, 2.7) -- (3, 1.9) node[above,  yshift=1.2ex]{$R$};
\draw[<->] (3.4, 2.7) -- (5.1,2.45) node[above, xshift=-6ex, yshift=1ex]{$d_{\text{min}}$};
\end{tikzpicture}%

%% file: Bilder/1EV_Konvergenzen_Integrationspunkte.tex
% This file was created by matlab2tikz.
%
%The latest updates can be retrieved from
%  http://www.mathworks.com/matlabcentral/fileexchange/22022-matlab2tikz-matlab2tikz
%where you can also make suggestions and rate matlab2tikz.
%
\definecolor{mycolor1}{rgb}{0.00000,0.44700,0.74100}%
\definecolor{mycolor2}{rgb}{0.85000,0.32500,0.09800}%
\definecolor{mycolor3}{rgb}{0.92900,0.69400,0.12500}%
\begin{tikzpicture}

\begin{axis}[%
width=0.9in,
height=1.25in,
at={(0in,0.478in)},
scale only axis,
xmin=0,
xmax=64,
xlabel style={font=\color{white!15!black}},
xlabel={$N_P$},
ymode=log,
ymin=1e-12,
ymax=1,
yminorticks=true,
ylabel style={font=\color{white!15!black}},
ylabel={$\Delta \Re\text{(}\omega\text{)}$},
axis background/.style={fill=white},
xmajorgrids,
ymajorgrids,
yminorgrids,
legend columns=-1,
legend style={at={(-0.3,-0.525)}, anchor=south west, legend cell align=left, align=left, draw=white!15!black}
]
\addplot [color=mycolor1, mark size=3.0pt, mark=asterisk, mark options={solid, mycolor1}]
  table[row sep=crcr]{%
2	6.29191232975266e-05\\
4	9.76591758758843e-06\\
8	3.72595639379009e-07\\
16	3.05328428177839e-11\\
32	8.46996872453887e-12\\
64	5.29728967294324e-12\\
};
\addlegendentry{Current  }
\addplot [color=mycolor2, mark size=3.0pt, mark=o, mark options={solid, mycolor2}]
  table[row sep=crcr]{%
2	1\\
4	1.27132815374309e-05\\
8	3.28151473025431e-07\\
16	7.28252188500838e-11\\
32	1.75133131812938e-12\\
64	5.96427776960276e-12\\
};
\addlegendentry{Dipole  }
\addplot [color=mycolor3, mark size=3.0pt, mark=triangle, mark options={solid, mycolor3}]
  table[row sep=crcr]{%
2	0.000143055370483558\\
4	6.73021361134076e-05\\
8	2.03870687041791e-05\\
16	2.67641624975618e-06\\
32	5.36100678578277e-08\\
64	1.49290918539635e-11\\
};
\addlegendentry{Plane wave}

\end{axis}

\begin{axis}[%
width=0.9in,
height=1.25in,
at={(1.62in,0.478in)},
scale only axis,
xmin=0,
xmax=64,
xlabel style={font=\color{white!15!black}},
xlabel={$N_P$},
ymode=log,
ymin=1e-10,
ymax=1,
yminorticks=true,
ylabel style={font=\color{white!15!black}},
ylabel={$\Delta\Im\text{(}\omega\text{)}$},
axis background/.style={fill=white},
xmajorgrids,
ymajorgrids,
yminorgrids,
legend style={at={(0.48,0.65)},anchor=south west, legend cell align=left, align=left, draw=white!15!black}
]
\addplot [color=mycolor1, mark size=3.0pt, mark=asterisk, mark options={solid, mycolor1}]
  table[row sep=crcr]{%
2	0.015129249359595\\
4	0.00713316603059789\\
8	0.0010243671463907\\
16	7.77107284511825e-06\\
32	1.7886897300086e-09\\
64	1.06752782172302e-08\\
};
%\addlegendentry{Current}

\addplot [color=mycolor2, mark size=3.0pt, mark=o, mark options={solid, mycolor2}]
  table[row sep=crcr]{%
2	1\\
4	0.0119228302514776\\
8	0.00107066024738404\\
16	6.89588428247214e-06\\
32	2.16449466738529e-08\\
64	1.40228678294726e-09\\
};
%\addlegendentry{Dipole}

\addplot [color=mycolor3, mark size=3.0pt, mark=triangle, mark options={solid, mycolor3}]
  table[row sep=crcr]{%
2	0.0469176191847576\\
4	0.0407811258670804\\
8	0.0207636050723332\\
16	0.00399983899797218\\
32	0.000126365685976941\\
64	4.1998223043583e-08\\
};
%\addlegendentry{Plane wave}

\end{axis}
\end{tikzpicture}%

%% file: Bilder/1EV_Konvergenzen_Radius_Kontur_4Punkte.tex
% This file was created by matlab2tikz.
%
%The latest updates can be retrieved from
%  http://www.mathworks.com/matlabcentral/fileexchange/22022-matlab2tikz-matlab2tikz
%where you can also make suggestions and rate matlab2tikz.
%
\definecolor{mycolor1}{rgb}{0.00000,0.44700,0.74100}%
\definecolor{mycolor2}{rgb}{0.85000,0.32500,0.09800}%
\definecolor{mycolor3}{rgb}{0.92900,0.69400,0.12500}%
\begin{tikzpicture}

\begin{axis}[%
width=0.9in,
height=1.25in,
at={(0.826in,0.473in)},
scale only axis,
xmode=log,
xmin=0.01,
xmax=1,
xminorticks=true,
xlabel style={font=\color{white!15!black}},
xlabel={$R/d_{\text{min}}$},
ymode=log,
ymin=1e-14,
ymax=0.01,
yminorticks=true,
ylabel style={font=\color{white!15!black}},
ylabel={$\Delta\Re\text{(}\omega\text{)}$},
axis background/.style={fill=white},
xmajorgrids,
xminorgrids,
ymajorgrids,
yminorgrids,
legend columns=-1,
legend style={at={(-0.3,-0.525)}, anchor=south west, legend cell align=left, align=left, draw=white!15!black}
]
\addplot [color=mycolor1,mark size=3.0pt, mark=asterisk, mark options={solid, mycolor1}]
  table[row sep=crcr]{%
0.01	9.56666357742835e-13\\
0.05	9.48038678187644e-10\\
0.1	1.52388452866017e-08\\
0.5	9.76591880467918e-06\\
0.9	0.000121219035596988\\
};
\addlegendentry{Current  }

\addplot [color=mycolor2,mark size=3.0pt, mark=o, mark options={solid, mycolor2}]
  table[row sep=crcr]{%
0.01	4.31930049885495e-12\\
0.05	1.22272257879623e-09\\
0.1	1.94906198065371e-08\\
0.5	1.27132821182177e-05\\
0.9	0.000161986883707657\\
};
\addlegendentry{Dipole  }

\addplot [color=mycolor3,mark size=3.0pt, mark=triangle, mark options={solid, mycolor3}]
  table[row sep=crcr]{%
0.01	6.17984624216835e-12\\
0.05	5.30165720981321e-09\\
0.1	8.49310472137157e-08\\
0.5	6.73021356222028e-05\\
0.9	0.000118647218397591\\
};
\addlegendentry{Plane wave}

\end{axis}

\begin{axis}[%
width=0.9in,
height=1.25in,
at={(2.4460in,0.478in)},
scale only axis,
xmode=log,
xmin=0.01,
xmax=1,
xminorticks=true,
xlabel style={font=\color{white!15!black}},
xlabel={$R/d_{\text{min}}$},
ymode=log,
ymin=1e-10,
ymax=100,
yminorticks=true,
ylabel style={font=\color{white!15!black}},
ylabel={$\Delta\Im\text{(}\omega\text{)}$},
axis background/.style={fill=white},
xmajorgrids,
xminorgrids,
ymajorgrids,
yminorgrids,
legend style={at={(0.06,0.65)}, anchor=south west, legend cell align=left, align=left, draw=white!15!black}
]
\addplot [color=mycolor1,mark size=3.0pt, mark=asterisk, mark options={solid, mycolor1}]
  table[row sep=crcr]{%
0.01	5.84215291715518e-09\\
0.05	6.34940188336005e-07\\
0.1	1.00737765465165e-05\\
0.5	0.00713315903240599\\
0.9	0.274821825494173\\
};

\addplot [color=mycolor2,mark size=3.0pt, mark=o, mark options={solid, mycolor2}]
  table[row sep=crcr]{%
0.01	3.40634558479544e-09\\
0.05	1.01494185236526e-06\\
0.1	1.64083636916165e-05\\
0.5	0.0119228329713644\\
0.9	0.748635009182997\\
};

\addplot [color=mycolor3,mark size=3.0pt, mark=triangle, mark options={solid, mycolor3}]
  table[row sep=crcr]{%
0.01	1.88017824848559e-10\\
0.05	3.00439820676722e-06\\
0.1	4.82369462773184e-05\\
0.5	0.0407811262387315\\
0.9	1.19825890730735\\
};

\end{axis}
\end{tikzpicture}%

%% file: Bilder/4EV_Integrationskontur.tex
% This file was created by matlab2tikz.
%
%The latest updates can be retrieved from
%  http://www.mathworks.com/matlabcentral/fileexchange/22022-matlab2tikz-matlab2tikz
%where you can also make suggestions and rate matlab2tikz.
%
\definecolor{mycolor1}{rgb}{0.00000,0.44700,0.74100}%
\definecolor{mycolor2}{rgb}{0.85000,0.32500,0.09800}%
\definecolor{mycolor3}{rgb}{0.92900,0.69400,0.12500}%
\definecolor{mycolor4}{rgb}{0.49400,0.18400,0.55600}%
\begin{tikzpicture}

\begin{axis}[%
width=2.25in,
height=1.35in,
at={(0.766in,0.486in)},
scale only axis,
xmin=1.921,
xmax=1.930,
xlabel style={font=\color{white!15!black}},
xlabel={$\Re\text{(}\omega\text{) }\cdot\text{ 10}^{\text{15}}\, [s^{-1}]$},
ymin=-3,
ymax=1,
ylabel style={font=\color{white!15!black}},
ylabel={$\Im\text{(}\omega\text{) }\cdot\text{ 10}^{\text{12}}\, [s^{-1}]$},
axis background/.style={fill=white},
xmajorgrids,
ymajorgrids,
legend style={at={(0.59,0.56)}, anchor=south west, legend cell align=left, align=left, draw=white!15!black}]%0, -0,625
\addplot [color=red, line width=2.0pt]
  table[row sep=crcr]{%
1.926	-0.3887\\
1.92599939763739	-0.364158771477088\\
1.92599759091241	-0.339632325672582\\
1.92599458091336	-0.315135436400332\\
1.92599036945334	-0.290682859670439\\
1.9259849590692	-0.266289324800784\\
1.92597835301993	-0.241969525544638\\
1.92597055528478	-0.217738111239699\\
1.92596157056081	-0.193609677983872\\
1.92595140426008	-0.16959875984313\\
1.92594006250639	-0.145719820096736\\
1.92592755213159	-0.121987242525102\\
1.92591388067146	-0.0984153227455378\\
1.92589905636119	-0.0750182596011087\\
1.92588308813037	-0.0518101466077798\\
1.92586598559767	-0.0288049634650118\\
1.92584775906502	-0.00601656763491025\\
1.92582841951141	0.0165413140049897\\
1.92580797858625	0.0388550934302819\\
1.92578644860239	0.0609113296546067\\
1.9257638425287	0.0826967368259977\\
1.92574017398222	0.104198192229784\\
1.92571545722	0.125402744193222\\
1.9256897071305	0.146297619887097\\
1.9256629392246	0.166870233019602\\
1.9256351696263	0.187108191417845\\
1.92560641506296	0.206999304492433\\
1.92557669285525	0.226531590580627\\
1.92554602090673	0.245693284163645\\
1.92551441769301	0.264472842953777\\
1.92548190225071	0.282858954847018\\
1.9254484941659	0.300840544737067\\
1.92541421356237	0.318406781186547\\
1.92537908108947	0.335547082951467\\
1.92534311790969	0.352251125354959\\
1.92530634568591	0.368508846506485\\
1.92526878656833	0.384310453362737\\
1.92523046318116	0.399646427626606\\
1.92519139860898	0.414507531480645\\
1.92515161638284	0.428884813151584\\
1.92511114046604	0.442769612302545\\
1.92506999523977	0.456153565249707\\
1.92502820548839	0.469028610000272\\
1.92498579638446	0.481386991108711\\
1.92494279347365	0.493221264348355\\
1.92489922265931	0.504524301195515\\
1.92485511018686	0.515289293123443\\
1.92481048262801	0.525509755703531\\
1.92476536686473	0.535179532511287\\
1.92471979007307	0.544292798834739\\
1.92467377970678	0.552844065183021\\
1.9246273634808	0.560828180593037\\
1.92458056935451	0.568240335732209\\
1.92453342551495	0.57507606579544\\
1.92448596035981	0.581331253194544\\
1.92443820248031	0.587002130038528\\
1.92439018064403	0.59208528040323\\
1.92434192377752	0.596577642388941\\
1.92429346094891	0.600476509964781\\
1.9242448213504	0.60377953459871\\
1.92419603428066	0.606484726672197\\
1.9241471291272	0.60859045667869\\
1.92409813534865	0.610095456205172\\
1.92404908245705	0.610998818696204\\
1.924	0.6113\\
1.92395091754295	0.610998818696204\\
1.92390186465134	0.610095456205172\\
1.9238528708728	0.60859045667869\\
1.92380396571934	0.606484726672197\\
1.9237551786496	0.60377953459871\\
1.92370653905109	0.600476509964781\\
1.92365807622248	0.596577642388941\\
1.92360981935597	0.59208528040323\\
1.92356179751969	0.587002130038528\\
1.92351403964019	0.581331253194544\\
1.92346657448505	0.57507606579544\\
1.92341943064549	0.568240335732209\\
1.9233726365192	0.560828180593037\\
1.92332622029322	0.552844065183021\\
1.92328020992693	0.544292798834739\\
1.92323463313527	0.535179532511287\\
1.92318951737199	0.525509755703531\\
1.92314488981314	0.515289293123443\\
1.92310077734069	0.504524301195515\\
1.92305720652635	0.493221264348355\\
1.92301420361554	0.481386991108711\\
1.92297179451161	0.469028610000272\\
1.92293000476023	0.456153565249707\\
1.92288885953396	0.442769612302545\\
1.92284838361716	0.428884813151584\\
1.92280860139101	0.414507531480645\\
1.92276953681884	0.399646427626606\\
1.92273121343167	0.384310453362737\\
1.92269365431409	0.368508846506484\\
1.92265688209031	0.352251125354959\\
1.92262091891053	0.335547082951467\\
1.92258578643763	0.318406781186547\\
1.9225515058341	0.300840544737067\\
1.92251809774929	0.282858954847018\\
1.92248558230699	0.264472842953777\\
1.92245397909327	0.245693284163645\\
1.92242330714475	0.226531590580627\\
1.92239358493704	0.206999304492433\\
1.9223648303737	0.187108191417845\\
1.92233706077539	0.166870233019602\\
1.9223102928695	0.146297619887097\\
1.92228454278	0.125402744193222\\
1.92225982601778	0.104198192229784\\
1.9222361574713	0.0826967368259978\\
1.92221355139761	0.0609113296546068\\
1.92219202141375	0.0388550934302819\\
1.92217158048859	0.0165413140049899\\
1.92215224093498	-0.00601656763491019\\
1.92213401440233	-0.0288049634650117\\
1.92211691186963	-0.0518101466077797\\
1.92210094363881	-0.0750182596011086\\
1.92208611932854	-0.0984153227455376\\
1.92207244786841	-0.121987242525102\\
1.92205993749361	-0.145719820096736\\
1.92204859573992	-0.16959875984313\\
1.92203842943919	-0.193609677983871\\
1.92202944471522	-0.217738111239699\\
1.92202164698007	-0.241969525544638\\
1.9220150409308	-0.266289324800784\\
1.92200963054666	-0.290682859670439\\
1.92200541908664	-0.315135436400332\\
1.92200240908759	-0.339632325672582\\
1.92200060236261	-0.364158771477088\\
1.922	-0.3887\\
1.92200060236261	-0.413241228522912\\
1.92200240908759	-0.437767674327418\\
1.92200541908664	-0.462264563599667\\
1.92200963054666	-0.486717140329561\\
1.9220150409308	-0.511110675199216\\
1.92202164698007	-0.535430474455362\\
1.92202944471522	-0.559661888760301\\
1.92203842943919	-0.583790322016128\\
1.92204859573992	-0.60780124015687\\
1.92205993749361	-0.631680179903264\\
1.92207244786841	-0.655412757474898\\
1.92208611932854	-0.678984677254462\\
1.92210094363881	-0.702381740398891\\
1.92211691186963	-0.72558985339222\\
1.92213401440233	-0.748595036534988\\
1.92215224093498	-0.77138343236509\\
1.92217158048859	-0.79394131400499\\
1.92219202141375	-0.816255093430282\\
1.92221355139761	-0.838311329654607\\
1.9222361574713	-0.860096736825998\\
1.92225982601778	-0.881598192229784\\
1.92228454278	-0.902802744193222\\
1.9223102928695	-0.923697619887097\\
1.92233706077539	-0.944270233019602\\
1.9223648303737	-0.964508191417845\\
1.92239358493704	-0.984399304492433\\
1.92242330714475	-1.00393159058063\\
1.92245397909327	-1.02309328416365\\
1.92248558230699	-1.04187284295378\\
1.92251809774929	-1.06025895484702\\
1.9225515058341	-1.07824054473707\\
1.92258578643763	-1.09580678118655\\
1.92262091891053	-1.11294708295147\\
1.92265688209031	-1.12965112535496\\
1.92269365431409	-1.14590884650648\\
1.92273121343167	-1.16171045336274\\
1.92276953681884	-1.17704642762661\\
1.92280860139101	-1.19190753148065\\
1.92284838361716	-1.20628481315158\\
1.92288885953396	-1.22016961230255\\
1.92293000476023	-1.23355356524971\\
1.92297179451161	-1.24642861000027\\
1.92301420361554	-1.25878699110871\\
1.92305720652635	-1.27062126434835\\
1.92310077734069	-1.28192430119552\\
1.92314488981314	-1.29268929312344\\
1.92318951737199	-1.30290975570353\\
1.92323463313527	-1.31257953251129\\
1.92328020992693	-1.32169279883474\\
1.92332622029322	-1.33024406518302\\
1.9233726365192	-1.33822818059304\\
1.92341943064549	-1.34564033573221\\
1.92346657448505	-1.35247606579544\\
1.92351403964019	-1.35873125319454\\
1.92356179751969	-1.36440213003853\\
1.92360981935597	-1.36948528040323\\
1.92365807622248	-1.37397764238894\\
1.92370653905109	-1.37787650996478\\
1.9237551786496	-1.38117953459871\\
1.92380396571934	-1.3838847266722\\
1.9238528708728	-1.38599045667869\\
1.92390186465134	-1.38749545620517\\
1.92395091754295	-1.3883988186962\\
1.924	-1.3887\\
1.92404908245705	-1.3883988186962\\
1.92409813534865	-1.38749545620517\\
1.9241471291272	-1.38599045667869\\
1.92419603428066	-1.3838847266722\\
1.9242448213504	-1.38117953459871\\
1.92429346094891	-1.37787650996478\\
1.92434192377752	-1.37397764238894\\
1.92439018064403	-1.36948528040323\\
1.92443820248031	-1.36440213003853\\
1.92448596035981	-1.35873125319454\\
1.92453342551495	-1.35247606579544\\
1.92458056935451	-1.34564033573221\\
1.9246273634808	-1.33822818059304\\
1.92467377970678	-1.33024406518302\\
1.92471979007307	-1.32169279883474\\
1.92476536686473	-1.31257953251129\\
1.92481048262801	-1.30290975570353\\
1.92485511018686	-1.29268929312344\\
1.92489922265931	-1.28192430119552\\
1.92494279347365	-1.27062126434835\\
1.92498579638446	-1.25878699110871\\
1.92502820548839	-1.24642861000027\\
1.92506999523977	-1.23355356524971\\
1.92511114046604	-1.22016961230255\\
1.92515161638284	-1.20628481315158\\
1.92519139860898	-1.19190753148065\\
1.92523046318116	-1.17704642762661\\
1.92526878656833	-1.16171045336274\\
1.92530634568591	-1.14590884650648\\
1.92534311790969	-1.12965112535496\\
1.92537908108947	-1.11294708295147\\
1.92541421356237	-1.09580678118655\\
1.9254484941659	-1.07824054473707\\
1.92548190225071	-1.06025895484702\\
1.92551441769301	-1.04187284295378\\
1.92554602090673	-1.02309328416365\\
1.92557669285525	-1.00393159058063\\
1.92560641506296	-0.984399304492433\\
1.9256351696263	-0.964508191417846\\
1.9256629392246	-0.944270233019602\\
1.9256897071305	-0.923697619887097\\
1.92571545722	-0.902802744193222\\
1.92574017398222	-0.881598192229784\\
1.9257638425287	-0.860096736825998\\
1.92578644860239	-0.838311329654607\\
1.92580797858625	-0.816255093430282\\
1.92582841951141	-0.793941314004989\\
1.92584775906502	-0.77138343236509\\
1.92586598559767	-0.748595036534988\\
1.92588308813037	-0.725589853392221\\
1.92589905636119	-0.702381740398891\\
1.92591388067146	-0.678984677254462\\
1.92592755213159	-0.655412757474899\\
1.92594006250639	-0.631680179903264\\
1.92595140426008	-0.60780124015687\\
1.92596157056081	-0.583790322016129\\
1.92597055528478	-0.559661888760302\\
1.92597835301993	-0.535430474455361\\
1.9259849590692	-0.511110675199217\\
1.92599036945334	-0.48671714032956\\
1.92599458091336	-0.462264563599668\\
1.92599759091241	-0.437767674327418\\
1.92599939763739	-0.413241228522912\\
1.926	-0.3887\\
};
\addlegendentry{Contour}

\addplot [color=teal, only marks, mark size=4.0pt, mark=asterisk, mark options={solid, teal}]
  table[row sep=crcr]{%
1.92282292056358	-0.388784029895033\\
1.92384638620298	-0.528227369059214\\
1.92443133648057	-0.688913018274329\\
1.92498223942906	-0.878120032959263\\
1.92611689426939	-1.34318523911825\\
1.92684635316195	-1.65436077590149\\
1.92897611973416	-2.3763252276256\\
1.92983644339417	-2.73852685433129\\
};
\addlegendentry{Arnoldi}

\addplot [color=blue, only marks, mark size=2.0pt, mark=*, mark options={solid, blue}]
  table[row sep=crcr]{%
1.92282292055895	-0.388784030080882\\
1.92384638287177	-0.528227561089942\\
1.9244313355732	-0.688911969136386\\
1.92498223877665	-0.878118622416168\\
};
\addlegendentry{Riesz}

\end{axis}
\draw[color=mycolor1] (3.1,3.8) node{1};
\draw[color=mycolor2] (3.75,3.75) node{2};
\draw[color=mycolor3] (4.15,3.65) node{3};
\draw[color=mycolor4] (4.5,3.4) node{4};

\draw (1.5,-2) node{\includegraphics[scale=0.08]{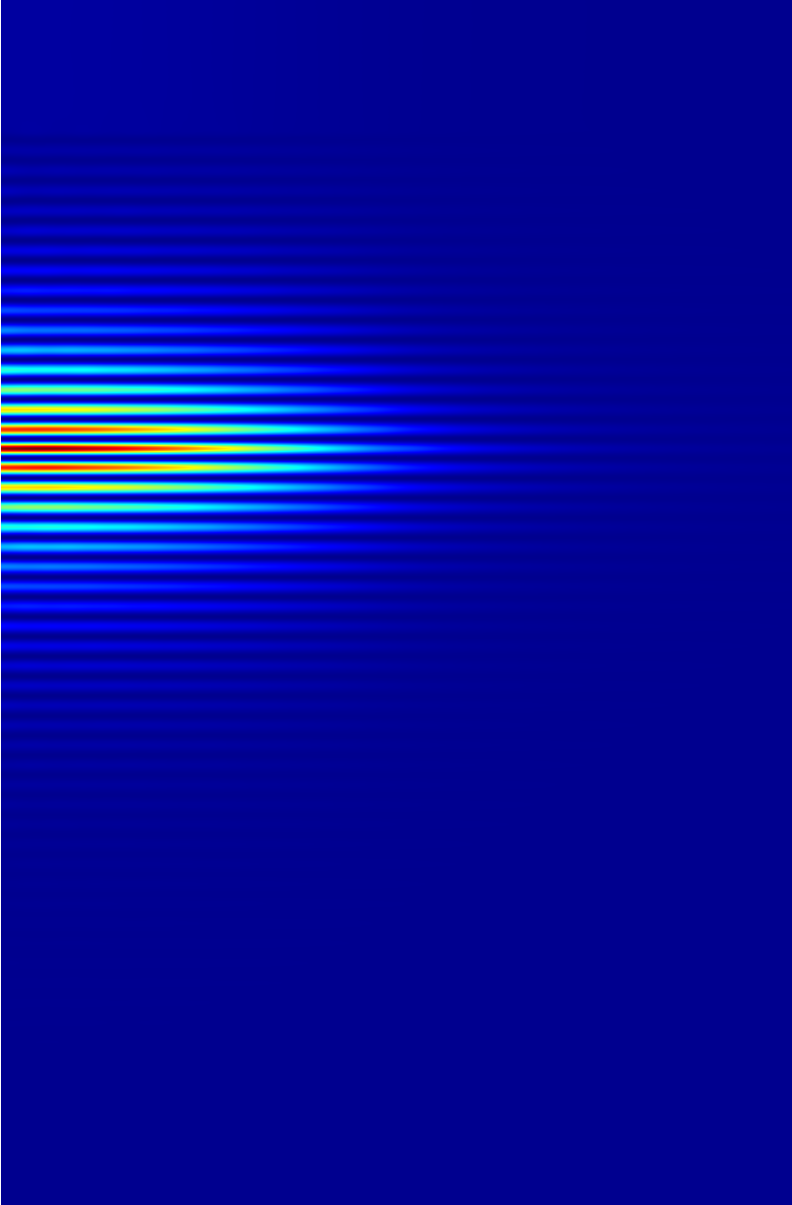}};
\draw (3.3,-2) node{\includegraphics[scale=0.08]{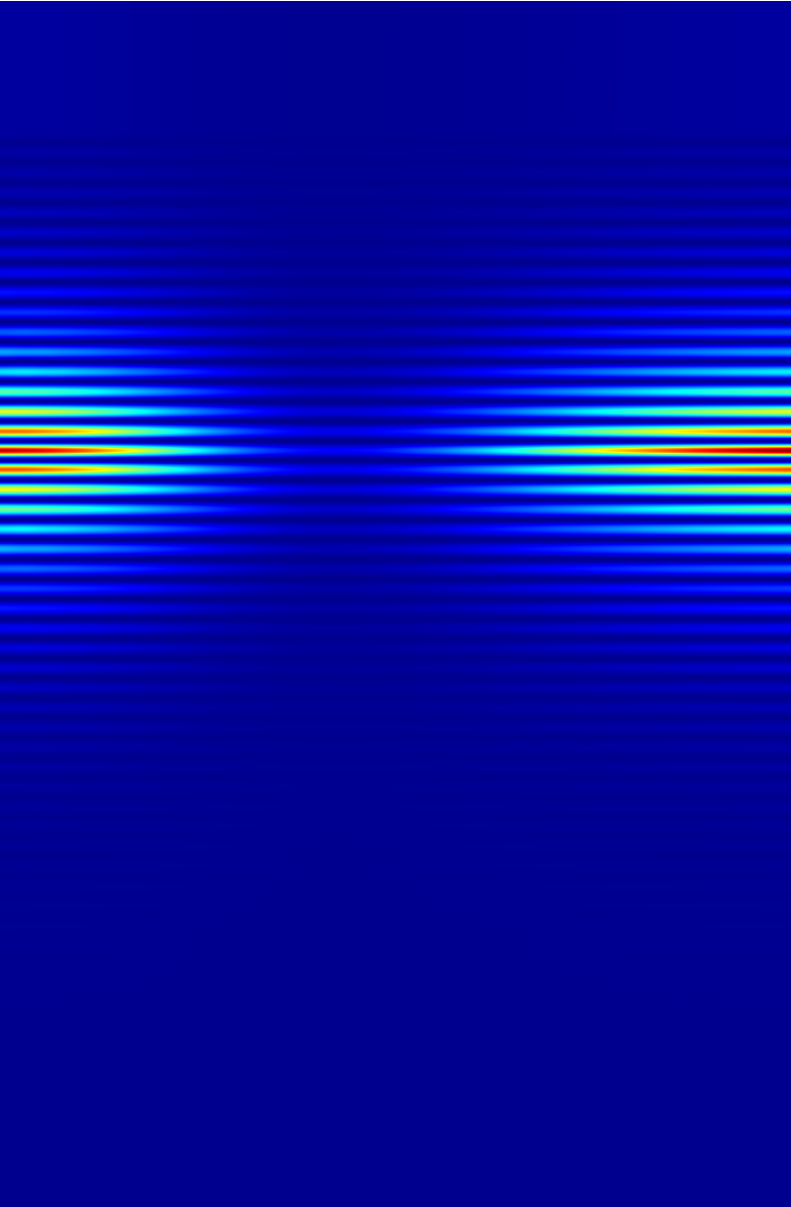}};
 \draw (5.1,-2) node{\includegraphics[scale=0.08]{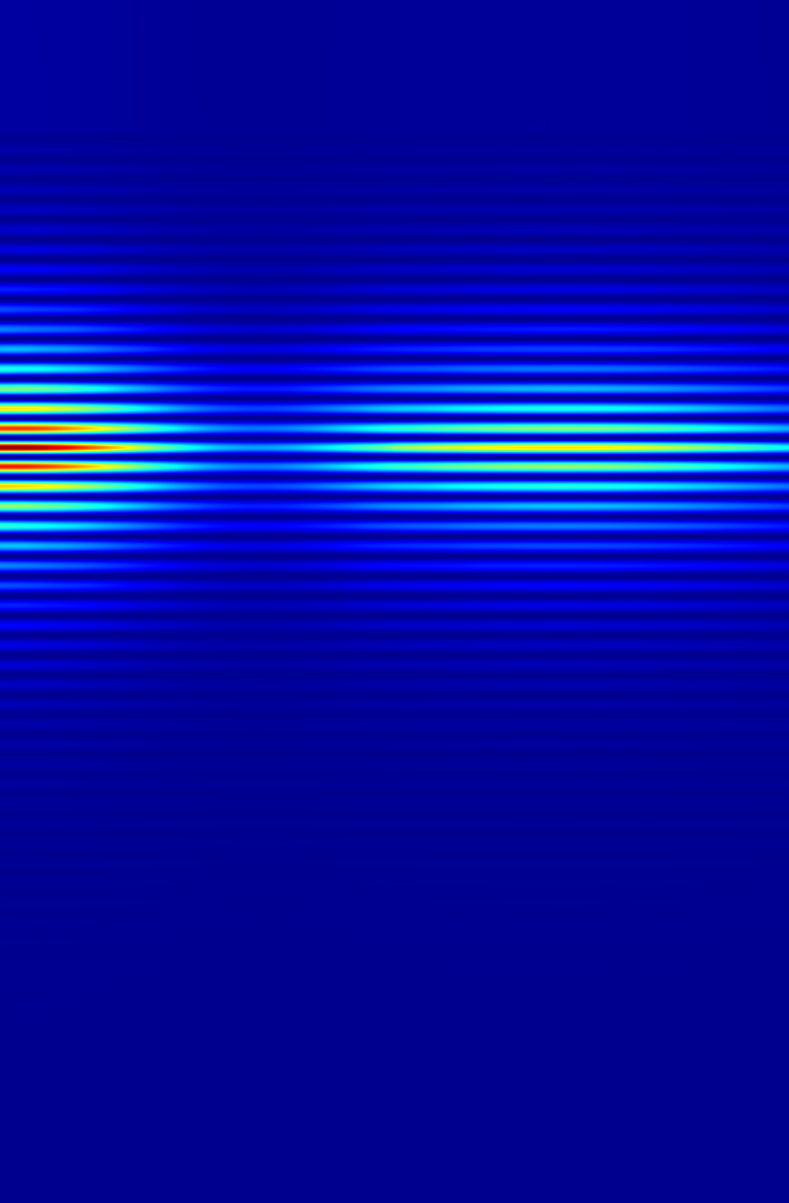}};
 \draw (7.0,-2) node{\includegraphics[scale=0.08]{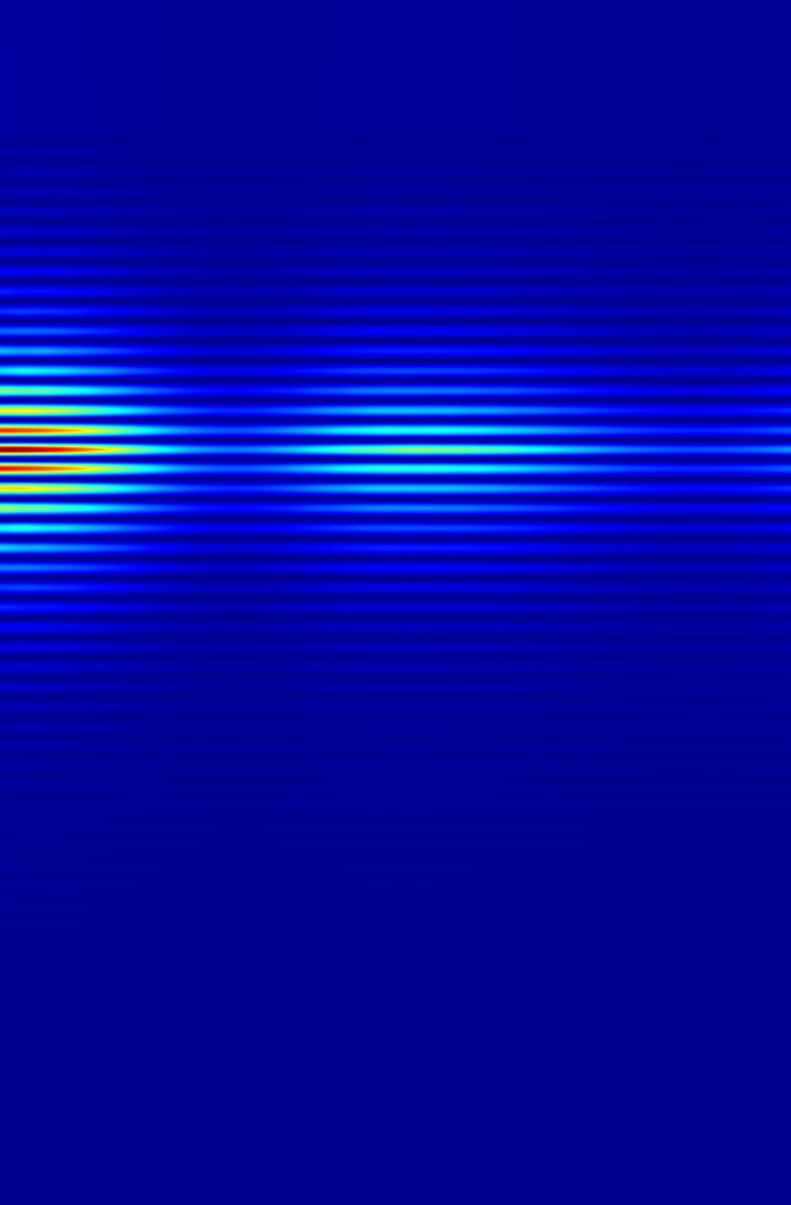}};

\draw[<-,semitransparent] (1.6,-0.5) -- (3.1,3.2);
\draw[<-,semitransparent] (3.3,-0.5) -- (3.75,3.1);
\draw[<-,semitransparent] (5.1,-0.5) -- (4.15,3.0);
\draw[<-,semitransparent] (7.1,-0.5) -- (4.55,2.9);
\end{tikzpicture}%

%% file: Bilder/4EV_Konvergenzen_Integrationspunke.tex
% This file was created by matlab2tikz.
%
%The latest updates can be retrieved from
%  http://www.mathworks.com/matlabcentral/fileexchange/22022-matlab2tikz-matlab2tikz
%where you can also make suggestions and rate matlab2tikz.
%
\definecolor{mycolor1}{rgb}{0.00000,0.44700,0.74100}%
\definecolor{mycolor2}{rgb}{0.85000,0.32500,0.09800}%
\definecolor{mycolor3}{rgb}{0.92900,0.69400,0.12500}%
\definecolor{mycolor4}{rgb}{0.49400,0.18400,0.55600}%

\begin{tikzpicture}

\begin{axis}[%
width=1.1in,
height=1.95in,
at={(1.489in,1.015in)},
scale only axis,
xlabel style={font=\color{white!15!black}},
xlabel={$N_P$},
xmax=300,
%xmode=log,
ymode=log,
ymin=1e-12,
ymax=0.01,
yminorticks=true,
ylabel style={font=\color{white!15!black}},
ylabel={$\Delta\Re\text{(}\omega\text{)}$},
axis background/.style={fill=white},
xmajorgrids,
ymajorgrids,
yminorgrids,
legend columns=3,
legend style={at={(-0.1,-0.625)}, anchor=south west, legend cell align=left, align=left, draw=white!15!black}
]
% Mode 1
\addplot [color=mycolor1, mark=asterisk, mark size=3.0pt, mark options={solid, mycolor1}]
  table[row sep=crcr]{%
8	1.6960088440147e-06\\
16	4.59993498642484e-07\\
32	1.0332714618451e-08\\
64	1.82915959779028e-11\\
128	9.08677019248288e-12\\
256	2.4073979722705e-12\\
};
\addlegendentry{1 - Current}
\addplot [color=mycolor1, dashed, mark=o, mark size=3.0pt, mark options={solid, mycolor1}]
  table[row sep=crcr]{%
8	9.72949304804764e-06\\
16	3.291234036593e-07\\
32	3.09093865921778e-08\\
64	4.6175468916283e-11\\
128	4.5147423130652e-11\\
256	2.49904707740409e-11\\
};
\addlegendentry{1 - Dipole}
\addplot [color=mycolor1, dotted, mark=triangle, mark size=3.0pt, mark options={solid, mycolor1}]
  table[row sep=crcr]{%
8	2.84891323633661e-05\\
16	1.19988110297425e-06\\
32	2.71482699559769e-07\\
64	2.75486691252225e-07\\
128	2.75474727852083e-07\\
256	2.75496335666072e-07\\
};
\addlegendentry{1 - Plane wave}
% Mode 2
\addplot [color=mycolor2, mark=asterisk, mark size=3.0pt, mark options={solid, mycolor2}]
  table[row sep=crcr]{%
8	3.89221370640659e-05\\
16	0.000100789551005655\\
32	2.70469637028547e-06\\
64	1.70772925196189e-09\\
128	1.30475568008015e-09\\
256	1.73153507675562e-09\\
};
\addlegendentry{2 - Current}
\addplot [color=mycolor2, dashed, mark=o, mark size=3.0pt, mark options={solid, mycolor2}]
  table[row sep=crcr]{%
8	0.000414989732163063\\
16	0.000117588844979192\\
32	2.3960355927106e-06\\
64	2.82387605838028e-09\\
128	2.71224266522571e-09\\
256	3.02030624257282e-09\\
};
\addlegendentry{2 - Dipole}
\addplot [color=mycolor2, dotted, mark=triangle, mark size=3.0pt, mark options={solid, mycolor2}]
  table[row sep=crcr]{%
8	0.000259167145590722\\
16	0.000142906946238765\\
32	0.000163434531552082\\
64	0.000163330480622037\\
128	0.000163330775342943\\
256	0.000163330011038417\\
};
\addlegendentry{2 - Plane wave}

% Mode 3
\addplot [color=mycolor3, mark=asterisk, mark size=3.0pt, mark options={solid, mycolor3}]
  table[row sep=crcr]{%
32	3.47610240591067e-06\\
64	3.89695769229941e-10\\
128	2.3321928483081e-10\\
256	4.71500324693013e-10\\
};
\addlegendentry{3 - Current}
\addplot [color=mycolor3, dashed, mark=o, mark size=3.0pt, mark options={solid, mycolor3}]
  table[row sep=crcr]{%
32	1.42330803979695e-06\\
64	4.35514629237311e-09\\
128	2.44323785986503e-09\\
256	2.30780070757013e-09\\
};
\addlegendentry{3 - Dipole}
\addplot [color=mycolor3, dotted, mark=triangle, mark size=3.0pt, mark options={solid, mycolor3}]
  table[row sep=crcr]{%
8	2.52746610590702e-05\\
16	3.53875090342499e-05\\
32	8.72308390476287e-05\\
64	8.67184940029865e-05\\
128	8.67203332906155e-05\\
256	8.67163331087934e-05\\
};
\addlegendentry{3 - Plane wave}

% Mode 4
\addplot [color=mycolor4, mark=asterisk, mark size=3.0pt, mark options={solid, mycolor4}]
  table[row sep=crcr]{%
8	0.000122536986693296\\
16	4.95244557104462e-05\\
32	6.17016296525561e-07\\
64	1.81078683667952e-10\\
128	2.53432468106664e-10\\
256	3.38920010084614e-10\\
};
\addlegendentry{4 - Current}
\addplot [color=mycolor4, dashed, mark=o, mark size=3.0pt, mark options={solid, mycolor4}]
  table[row sep=crcr]{%
16	4.05812898437024e-05\\
32	1.33993003918053e-07\\
64	6.81253817899616e-10\\
128	5.63561433336528e-10\\
256	6.16402829956462e-10\\
};
\addlegendentry{4 - Dipole}
\addplot [color=mycolor4, dotted, mark=triangle, mark size=3.0pt, mark options={solid, mycolor4}]
  table[row sep=crcr]{%
16	6.63169428265026e-05\\
32	1.26996376522418e-05\\
64	1.24661705487306e-05\\
128	1.24667858760961e-05\\
256	1.24654820306898e-05\\
};
\addlegendentry{4 - Plane wave}
\end{axis}

\begin{axis}[%
width=1.1in,
height=1.95in,
at={(3.332in,1.015in)}, 
scale only axis,
xlabel style={font=\color{white!15!black}},
xlabel={$N_P$},
xmin=0,
xmax=300,
%xmode=log,
ymode=log,
ymin=1e-10,
ymax=100,
%yminorticks=true,
ylabel style={font=\color{white!15!black}},
ylabel={$\Delta\Im\text{(}\omega\text{)}$},
axis background/.style={fill=white},
xmajorgrids,
ymajorgrids,
yminorgrids
]
\addplot [color=mycolor1, mark=asterisk, mark size=3.0pt, mark options={solid, mycolor1}, forget plot]
  table[row sep=crcr]{%
8	0.00964876804033256\\
16	0.00311440516309337\\
32	4.9989545182658e-05\\
64	5.2087995590834e-08\\
128	1.65587990653755e-09\\
256	4.7818990329008e-10\\
};
\addplot [color=mycolor2, mark=asterisk, mark size=3.0pt, mark options={solid, mycolor2}, forget plot]
  table[row sep=crcr]{%
8	0.455364027092749\\
16	0.061393196038282\\
32	0.000949883300007078\\
64	1.34459532369906e-05\\
128	1.00873829469513e-06\\
256	3.63535152148577e-07\\
};
\addplot [color=mycolor3, mark=asterisk, mark size=3.0pt, mark options={solid, mycolor3}, forget plot]
  table[row sep=crcr]{%
32	0.000261929659693029\\
64	1.63717516177345e-05\\
128	3.25705051909859e-06\\
256	1.52288883504577e-06\\
};
\addplot [color=mycolor4, mark=asterisk, mark size=3.0pt, mark options={solid, mycolor4}, forget plot]
  table[row sep=crcr]{%
8	0.0145130321871706\\
16	0.081362244058938\\
32	0.000241108996568411\\
64	3.38258499824509e-06\\
128	1.92517908439899e-06\\
256	1.60632150734242e-06\\
};
\addplot [color=mycolor1, dashed, mark=o, mark size=3.0pt, mark options={solid, mycolor1}, forget plot]
  table[row sep=crcr]{%
8	0.0638317004472113\\
16	0.00408488319099764\\
32	7.22533030377593e-05\\
64	3.58602780616604e-07\\
128	1.72805691210406e-07\\
256	3.89069413808521e-08\\
};
\addplot [color=mycolor2, dashed, mark=o, mark size=3.0pt, mark options={solid, mycolor2}, forget plot]
  table[row sep=crcr]{%
8	0.620321178230392\\
16	0.181721354646085\\
32	0.0248271216511959\\
64	6.12741148854542e-05\\
128	4.22818960532806e-05\\
256	1.52836733176496e-05\\
};
\addplot [color=mycolor3, dashed, mark=o, mark size=3.0pt, mark options={solid, mycolor3}, forget plot]
  table[row sep=crcr]{%
32	0.0200597699868155\\
64	4.43749960267946e-05\\
128	3.16798844452622e-05\\
256	1.0658438476586e-05\\
};
\addplot [color=mycolor4, dashed, mark=o, mark size=3.0pt, mark options={solid, mycolor4}, forget plot]
  table[row sep=crcr]{%
16	0.0101427826043022\\
32	0.00202058299443108\\
64	2.48731512912129e-06\\
128	1.56852410076208e-06\\
256	3.79367158786814e-07\\
};
\addplot [color=mycolor1, dotted, mark=triangle, mark size=3.0pt, mark options={solid, rotate=180, mycolor1}, forget plot]
  table[row sep=crcr]{%
%4	0.00021613515106468\\
8	0.161191896626659\\
16	0.00546051764511074\\
32	0.0030195959460672\\
64	0.0030338682715137\\
128	0.00303386022461884\\
256	0.00303399215774743\\
};
\addplot [color=mycolor2, dotted, mark=triangle, mark size=3.0pt, mark options={solid, mycolor2}, forget plot]
  table[row sep=crcr]{%
%4	0.264142634842483\\
8	1.34120641843087\\
16	0.24512235178561\\
32	0.242112266950356\\
64	0.242112765374077\\
128	0.242112379997794\\
256	0.242111492015596\\
};
\addplot [color=mycolor3, dotted, mark=triangle, mark size=3.0pt, mark options={solid, mycolor3}, forget plot]
  table[row sep=crcr]{%
8	0.223802724889556\\
16	0.187551342968369\\
32	0.0951299494240697\\
64	0.0953408005291272\\
128	0.0953401291894801\\
256	0.0953363054854987\\
};
\addplot [color=mycolor4, dotted, mark=triangle, mark size=3.0pt, mark options={solid, mycolor4}, forget plot]
  table[row sep=crcr]{%
16	0.110640641910403\\
32	0.0146468881477758\\
64	0.0146700195247742\\
128	0.0146711781532267\\
256	0.0146682893297315\\
};
\end{axis}
\end{tikzpicture}%

%% file: Bilder/Konvergenz_4EV_N_Expected_Eigenvalues.tex
% This file was created by matlab2tikz.
%
%The latest updates can be retrieved from
%  http://www.mathworks.com/matlabcentral/fileexchange/22022-matlab2tikz-matlab2tikz
%where you can also make suggestions and rate matlab2tikz.
\definecolor{mycolor1}{rgb}{0.00000,0.44700,0.74100}%
\definecolor{mycolor2}{rgb}{0.85000,0.32500,0.09800}%
\definecolor{mycolor3}{rgb}{0.92900,0.69400,0.12500}%
\definecolor{mycolor4}{rgb}{0.49400,0.18400,0.55600}%
\begin{tikzpicture}

\begin{axis}[%
width=2.5in,
height=1in,
at={(1.489in,1.015in)},
scale only axis,
xmin=1,
xmax=6,
xlabel style={font=\color{white!15!black}},
xlabel={$m$},
ymode=log,
ymin=1.82915959779028e-11,
ymax=1,
yminorticks=true,
ylabel style={font=\color{white!15!black}},
ylabel={$\Delta\Re\text{(}\omega\text{)}$},
axis background/.style={fill=white},
xmajorgrids,
ymajorgrids,
yminorgrids,
legend columns=3,
legend style={at={(0.5,-1.2)}, anchor=south west, legend cell align=left, align=left, draw=white!15!black}
]
% %%%%%%%%%%%%%%%%% Mode 1 %%%%%%%%%%%%%%%%%%%%%%%%%%%%%%%%%%%%%%%%%%%%%%%%%%%%%%%%%%%%%%%%%%%%
\addplot [color=mycolor1, mark=asterisk, mark size=3.0pt, mark options={solid, mycolor1}]
  table[row sep=crcr]{%
1	1\\
2	3.06882367572386e-05\\
3	1.47506022534238e-06\\
4	1.82915959779028e-11\\
5	1.82915959779028e-11\\
6	1.82915959779028e-11\\
%7	1.82915959779028e-11\\
%8	1.82915959779028e-11\\
%9	1.82915959779028e-11\\
};
\addlegendentry{1-Current}
\addplot [color=mycolor1, dashed,mark=o, mark size=3.0pt, mark options={solid, mycolor1}]
  table[row sep=crcr]{%
1	0.000482536393310624\\
2	2.901306815315e-05\\
3	1.47548297774007e-06\\
4	1.0400775748054e-10\\
5	1.0400775748054e-10\\
6	1.0400775748054e-10\\
%7	1.0400775748054e-10\\
%8	1.0400775748054e-10\\
%9	1.0400775748054e-10\\
};
\addlegendentry{1-Dipole}
\addplot [color=mycolor1,dotted, mark=triangle, mark size=3.0pt, mark options={solid, mycolor1}]
  table[row sep=crcr]{%
1	1\\
2	6.50376729655356e-05\\
3	5.52194684424604e-06\\
4	2.75488438683028e-07\\
5	1.03871369466233e-10\\
6	1.03871369466233e-10\\
%7	1.03871369466233e-10\\
%8	1.03871369466233e-10\\
%9	1.03871369466233e-10\\
};
\addlegendentry{1-Plane wave   }
% %%%%%%%%%%%%%%%%% Mode 2 %%%%%%%%%%%%%%%%%%%%%%%%%%%%%%%%%%%%%%%%%%%%%%%%%%%%%%%%%%%%%%%%%%%%
\addplot [color=mycolor2, mark=asterisk, mark size=3.0pt, mark options={solid, mycolor2}]
  table[row sep=crcr]{%
1	1\\
2	1\\
3	1\\
4	1.7077344498821e-09\\
5	1.7077345798301e-09\\
6	1.7077344498821e-09\\
%7	1.7077344498821e-09\\
%8	1.7077344498821e-09\\
%9	1.7077344498821e-09\\
};
\addlegendentry{2-Current}
\addplot [color=mycolor2,dashed, mark=o, mark size=3.0pt, mark options={solid, mycolor2}]
  table[row sep=crcr]{%
1	1\\
2	1\\
3	1\\
4	1.19669705258739e-08\\
5	1.19669705258739e-08\\
6	1.19669705258739e-08\\
%7	1.19669705258739e-08\\
%8	1.19669705258739e-08\\
%9	1.19669705258739e-08\\
};
\addlegendentry{2-Dipole}
\addplot [color=mycolor2,dotted, mark=triangle, mark size=3.0pt, mark options={solid, mycolor2}]
  table[row sep=crcr]{%
1	1\\
2	0.000208410338461502\\
3	9.07707385534345e-05\\
4	0.000163330451543571\\
5	1.05312261910824e-07\\
6	1.05312262040772e-07\\
%7	1.05312261910824e-07\\
%8	1.05312262040772e-07\\
%9	1.05312262040772e-07\\
};
\addlegendentry{2-Plane wave   }

% %%%%%%%%%%%%%%%%% Mode 3 %%%%%%%%%%%%%%%%%%%%%%%%%%%%%%%%%%%%%%%%%%%%%%%%%%%%%%%%%%%%%%%%%%%%
\addplot [color=mycolor3, mark=asterisk, mark size=3.0pt, mark options={solid, mycolor3}]
  table[row sep=crcr]{%
1	1\\
2	1\\
3	0.000151362272758174\\
4	3.89675633411496e-10\\
5	3.8967550350299e-10\\
6	3.8967550350299e-10\\
%7	3.8967550350299e-10\\
%8	3.8967550350299e-10\\
%9	3.8967550350299e-10\\
};
\addlegendentry{3-Current}
\addplot [color=mycolor3,dashed, mark=o, mark size=3.0pt, mark options={solid, mycolor3}]
  table[row sep=crcr]{%
1	1\\
2	1\\
3	0.000134590680038723\\
4	1.56408626690973e-08\\
5	1.56408626690973e-08\\
6	1.56408626690973e-08\\
%7	1.56408626690973e-08\\
%8	1.56408626690973e-08\\
%9	1.56408626690973e-08\\
};
\addlegendentry{3-Dipole}
\addplot [color=mycolor3,dotted, mark=triangle, mark size=3.0pt, mark options={solid, mycolor3}]
  table[row sep=crcr]{%
1	1\\
2	1\\
3	1\\
4	8.6718444150987e-05\\
5	6.20940234316363e-08\\
6	6.20940234316363e-08\\
%7	6.20940234316363e-08\\
%8	6.20940234316363e-08\\
%9	6.20940234316363e-08\\
};
\addlegendentry{3-Plane wave   }
% %%%%%%%%%%%%%%%%% Mode 4 %%%%%%%%%%%%%%%%%%%%%%%%%%%%%%%%%%%%%%%%%%%%%%%%%%%%%%%%%%%%%%%%%%%%
\addplot [color=mycolor4, mark=asterisk, mark size=3.0pt, mark options={solid, mycolor4}]
  table[row sep=crcr]{%
1	1\\
2	0.000116998818656399\\
3	1.98507710509234e-05\\
4	1.81076995340686e-10\\
5	1.81076995340686e-10\\
6	1.81076995340686e-10\\
%7	1.81076995340686e-10\\
%8	1.81076995340686e-10\\
%9	1.81076995340686e-10\\
};
\addlegendentry{4-Current}
\addplot [color=mycolor4,dashed, mark=o, mark size=3.0pt, mark options={solid, mycolor4}]
  table[row sep=crcr]{%
1	1\\
2	9.77573043959947e-05\\
3	1.53490598748399e-05\\
4	5.91533379724963e-10\\
5	5.91533379724963e-10\\
6	5.91533379724963e-10\\
%7	5.91533379724963e-10\\
%8	5.91533379724963e-10\\
%9	5.91533379724963e-10\\
};
\addlegendentry{4-Dipole}
\addplot [color=mycolor4,dotted, mark=triangle, mark size=3.0pt, mark options={solid, mycolor4}]
  table[row sep=crcr]{%
1	1\\
2	1\\
3	8.31728222043164e-05\\
4	1.24661168042088e-05\\
5	8.76656867494278e-09\\
6	8.76656867494278e-09\\
%7	8.76656867494278e-09\\
%8	8.76656867494278e-09\\
%9	8.76656867494278e-09\\
};
\addlegendentry{4-Plane wave   }

\end{axis}

\begin{axis}[%
width=2.5in,
height=1in,
at={(5in,1.015in)}, 
scale only axis,
xmin=1,
xmax=6,
xlabel style={font=\color{white!15!black}},
xlabel={$m$},
ymode=log,
ymin=5.20887800692412e-08,
ymax=1.04318886520966,
yminorticks=true,
ylabel style={font=\color{white!15!black}},
ylabel={$\Delta\Im\text{(}\omega\text{)}$},
axis background/.style={fill=white},
xmajorgrids,
ymajorgrids,
yminorgrids
]
% %%%%%%%%%%%%%%%%% Mode 1 %%%%%%%%%%%%%%%%%%%%%%%%%%%%%%%%%%%%%%%%%%%%%%%%%%%%%%%%%%%%%%%%%%%%
\addplot [color=mycolor1, mark=asterisk, mark size=3.0pt, mark options={solid, mycolor1}, forget plot]
  table[row sep=crcr]{%
1	1\\
2	0.129727458647734\\
3	0.00963224532414004\\
4	5.20887813251602e-08\\
5	5.20887800692412e-08\\
6	5.20887813251602e-08\\
%7	5.20887810111805e-08\\
%8	5.2088780383221e-08\\
%9	5.20887810111805e-08\\
};
\addplot [color=mycolor1,dashed, mark=o, mark size=3.0pt, mark options={solid, mycolor1}, forget plot]
  table[row sep=crcr]{%
1	1.04318886520966\\
2	0.126060146770313\\
3	0.00939635338638872\\
4	1.71804882484246e-07\\
5	1.71804882641236e-07\\
6	1.71804883426185e-07\\
%7	1.71804883269195e-07\\
%8	1.71804884054145e-07\\
%9	1.71804884211134e-07\\
};
\addplot [color=mycolor1,dotted, mark=triangle, mark size=3.0pt, mark options={solid, mycolor1}, forget plot]
  table[row sep=crcr]{%
1	1\\
2	0.128648753644773\\
3	0.0273078817488994\\
4	0.00303386686216156\\
5	4.95583914864568e-07\\
6	4.95583916591457e-07\\
%7	4.95583915335538e-07\\
%8	4.95583915335538e-07\\
%9	4.95583916120487e-07\\
};
% %%%%%%%%%%%%%%%%% Mode 2 %%%%%%%%%%%%%%%%%%%%%%%%%%%%%%%%%%%%%%%%%%%%%%%%%%%%%%%%%%%%%%%%%%%%
\addplot [color=mycolor2, mark=asterisk, mark size=3.0pt, mark options={solid, mycolor2}, forget plot]
  table[row sep=crcr]{%
1	1\\
2	1\\
3	1\\
4	1.34460017055505e-05\\
5	1.34460016662645e-05\\
6	1.34460017001198e-05\\
%7	1.34460016742372e-05\\
%8	1.34460016086065e-05\\
%9	1.34460016780503e-05\\
};
\addplot [color=mycolor2,dashed, mark=o, mark size=3.0pt, mark options={solid, mycolor2}, forget plot]
  table[row sep=crcr]{%
1	1\\
2	1\\
3	1\\
4	0.000130667151027689\\
5	0.000130667151074139\\
6	0.000130667151011282\\
%7	0.000130667151071597\\
%8	0.000130667151074139\\
%9	0.000130667151106724\\
};
\addplot [color=mycolor2,dotted, mark=triangle, mark size=3.0pt, mark options={solid, mycolor2}, forget plot]
  table[row sep=crcr]{%
1	1\\
2	0.107766815969797\\
3	0.137995491923335\\
4	0.242112872160808\\
5	0.000107892196712945\\
6	0.000107892195767885\\
%7	0.000107892196223372\\
%8	0.000107892196217942\\
%9	0.000107892195803012\\
};
% %%%%%%%%%%%%%%%%% Mode 3 %%%%%%%%%%%%%%%%%%%%%%%%%%%%%%%%%%%%%%%%%%%%%%%%%%%%%%%%%%%%%%%%%%%%
\addplot [color=mycolor3, mark=asterisk, mark size=3.0pt, mark options={solid, mycolor3}, forget plot]
  table[row sep=crcr]{%
1	1\\
2	1\\
3	0.138669477262537\\
4	1.63717719410135e-05\\
5	1.63717719564293e-05\\
6	1.63717719427854e-05\\
%7	1.63717719263065e-05\\
%8	1.63717719096504e-05\\
%9	1.63717719238258e-05\\
};
\addplot [color=mycolor3,dashed, mark=o, mark size=3.0pt, mark options={solid, mycolor3}, forget plot]
  table[row sep=crcr]{%
1	1\\
2	1\\
3	0.0846186806050166\\
4	9.60313464860955e-05\\
5	9.60313465231287e-05\\
6	9.60313464813113e-05\\
%7	9.60313465665409e-05\\
%8	9.60313465247235e-05\\
%9	9.6031346539962e-05\\
};
\addplot [color=mycolor3,dotted, mark=triangle, mark size=3.0pt, mark options={solid, mycolor3}, forget plot]
  table[row sep=crcr]{%
1	1\\
2	1\\
3	1\\
4	0.095340764934722\\
5	2.34962236805055e-06\\
6	2.34962207302482e-06\\
%7	2.34962233739623e-06\\
%8	2.34962230780506e-06\\
%9	2.34962221619647e-06\\
};
% %%%%%%%%%%%%%%%%% Mode 4 %%%%%%%%%%%%%%%%%%%%%%%%%%%%%%%%%%%%%%%%%%%%%%%%%%%%%%%%%%%%%%%%%%%%
\addplot [color=mycolor4, mark=asterisk, mark size=3.0pt, mark options={solid, mycolor4}, forget plot]
  table[row sep=crcr]{%
1	1\\
2	0.155416082864684\\
3	0.0269344439600363\\
4	3.38258991152926e-06\\
5	3.38258991361446e-06\\
6	3.38258991208531e-06\\
%7	3.38258991125123e-06\\
%8	3.38258991917499e-06\\
%9	3.38258991139025e-06\\
};

\addplot [color=mycolor4,dashed, mark=o, mark size=3.0pt, mark options={solid, mycolor4}, forget plot]
  table[row sep=crcr]{%
1	1\\
2	0.0292121258537815\\
3	6.55419820159753e-05\\
4	2.060039392432e-05\\
5	2.06003939264052e-05\\
6	2.06003939251541e-05\\
%7	2.06003939286294e-05\\
%8	2.06003939283514e-05\\
%9	2.06003939287684e-05\\
};
\addplot [color=mycolor4,dotted, mark=triangle, mark size=3.0pt, mark options={solid, mycolor4}, forget plot]
  table[row sep=crcr]{%
1	1\\
2	1\\
3	0.0916915696374945\\
4	0.0146700766889264\\
5	5.81962339126604e-06\\
6	5.81962341531534e-06\\
%7	5.81962339627052e-06\\
%8	5.81962339724361e-06\\
%9	5.81962340197006e-06\\
};
\end{axis}

\end{tikzpicture}%

%% file: Applying_a_Riesz_projection_based_contour_integral_eigenvalue_solver_to_compute_resonance_modes_of_a_VCSEL.bbl
\begin{thebibliography}{10}
	
	\bibitem{Liu:19}
	Liu, A., Wolf, P., Lott, J.~A., and Bimberg, D., ``Vertical-cavity
	surface-emitting lasers for data communication and sensing,'' {\em Photon.
		Res.}~{\bf 7},  121 (2019).
	
	\bibitem{gkebski202030}
	G{{e}}bski, M., Wong, P.-S., Riaziat, M., and Lott, J.~A., ``30 {GHz} bandwidth
	temperature stable 980 nm vertical-cavity surface-emitting lasers with
	{AlAs}/{GaAs} bottom distributed {B}ragg reflectors for optical data
	communication,'' {\em J. Phys. Photonics}~{\bf 2},  035008 (2020).
	
	\bibitem{Moughames:20}
	Moughames, J., Porte, X., Thiel, M., Ulliac, G., Larger, L., Jacquot, M.,
	Kadic, M., and Brunner, D., ``Three-dimensional waveguide interconnects for
	scalable integration of photonic neural networks,'' {\em Optica}~{\bf 7},
	640 (2020).
	
	\bibitem{moench2016vcsel}
	Moench, H., Carpaij, M., Gerlach, P., Gronenborn, S., Gudde, R., Hellmig, J.,
	Kolb, J., and van~der Lee, A., ``{VCSEL}-based sensors for distance and
	velocity,'' {\em Proc. SPIE}~{\bf 9766},  97660A (2016).
	
	\bibitem{Haghighi2021}
	Haghighi, N. and Lott, J.~A., ``Electrically parallel three-element 980 nm
	{VCSEL} arrays with ternary and binary bottom {DBR} mirror layers,'' {\em
		Materials}~{\bf 14},  397 (2021).
	
	\bibitem{chen2022deep}
	Chen, Z., Sludds, A., Davis, R., Christen, I., Bernstein, L., Heuser, T.,
	Heermeier, N., Lott, J.~A., Reitzenstein, S., Hamerly, R., and Englund, D.,
	``Deep learning with coherent {VCSEL} neural networks,'' {\em arXiv} ,
	2207.05329 (2022).
	
	\bibitem{Bienstman}
	Bienstman, P., Baets, R., Vukusic, J., Larsson, A., Noble, M., Brunner, M.,
	Gulden, K., Debernardi, P., Fratta, L., Bava, G., Wenzel, H., Klein, B.,
	Conradi, O., Pregla, R., Riyopoulos, S., Seurin, J.-F., and Chuang, S.~L.,
	``Comparison of optical {VCSEL} models on the simulation of oxide-confined
	devices,'' {\em IEEE J. Quantum Electron.}~{\bf 37},  1631 (2001).
	
	\bibitem{saad2011numerical}
	Saad, Y.,  [{\em Numerical methods for large eigenvalue problems, Revised
		Edition}{\nolinebreak\hspace{0.1em}]}, SIAM (2011).
	
	\bibitem{Asakura_JSIAM_2009}
	Asakura, J., Sakurai, T., Tadano, H., Ikegami, T., and Kimura, K., ``{A
		numerical method for nonlinear eigenvalue problems using contour
		integrals},'' {\em JSIAM Lett.}~{\bf 1},  52 (2009).
	
	\bibitem{beyn2012integral}
	Beyn, W.-J., ``An integral method for solving nonlinear eigenvalue problems,''
	{\em Linear Algebra Its Appl}~{\bf 436},  3839 (2012).
	
	\bibitem{Gavin_JCompPhy_2018}
	Gavin, B., Miedlar, A., and Polizzi, E., ``{FEAST eigensolver for nonlinear
		eigenvalue problems},'' {\em J. Comput. Sci.}~{\bf 27},  107 (2018).
	
	\bibitem{binkowski2020riesz}
	Binkowski, F., Zschiedrich, L., and Burger, S., ``A {R}iesz-projection-based
	method for nonlinear eigenvalue problems,'' {\em J. Comput. Phys.}~{\bf 419},
	109678 (2020).
	
	\bibitem{betz2021rpexpand}
	Betz, F., Binkowski, F., and Burger, S., ``{RPExpand}: {S}oftware for {R}iesz
	projection expansion of resonance phenomena,'' {\em SoftwareX}~{\bf 15},
	100763 (2021).
	
	\bibitem{binkowski2022computation}
	Binkowski, F., Betz, F., Hammerschmidt, M., Schneider, P.-I., Zschiedrich, L.,
	and Burger, S., ``Computation of eigenfrequency sensitivities using {R}iesz
	projections for efficient optimization of nanophotonic resonators,'' {\em
		Commun. Phys.}~{\bf 5},  202 (2022).
	
	\bibitem{JCMsuite}
	{JCMwave GmbH}, ``{JCMsuite}.'' \url{https://jcmwave.com/}.
	\newblock Accessed: 2022-12-13.
	
	\bibitem{pomplun2007adaptive}
	Pomplun, J., Burger, S., Zschiedrich, L., and Schmidt, F., ``Adaptive finite
	element method for simulation of optical nano structures,'' {\em Phys. Status
		Solidi B}~{\bf 244},  3419 (2007).
	
	\bibitem{berenger1994perfectly}
	Berenger, J.-P., ``A perfectly matched layer for the absorption of
	electromagnetic waves,'' {\em J. Comput. Phys.}~{\bf 114},  185 (1994).
	
	\bibitem{zschiedrich2018riesz}
	Zschiedrich, L., Binkowski, F., Nikolay, N., Benson, O., Kewes, G., and Burger,
	S., ``Riesz-projection-based theory of light-matter interaction in dispersive
	nanoresonators,'' {\em Phys. Rev. A}~{\bf 98},  043806 (2018).
	
	\bibitem{Schaedle_jcp_2007}
	Sch\"adle, A., Zschiedrich, L., Burger, S., Klose, R., and Schmidt, F.,
	``Domain decomposition method for {M}axwell's equations: {S}cattering off
	periodic structures,'' {\em J. Comp. Phys.}~{\bf 226},  447 (2007).
	
	\bibitem{Gebski:19}
	G{{e}}bski, M., Lott, J.~A., and Czyszanowski, T., ``Electrically injected
	{VCSEL} with a composite {DBR} and {MHCG} reflector,'' {\em Opt.
		Express}~{\bf 27},  7139 (2019).
	
\end{thebibliography}
